\documentclass[fleqn, usenatbib]{mnras}

\usepackage{newtxtext, newtxmath}

\usepackage[T1]{fontenc}
\usepackage{ae, aecompl}

\usepackage{amsmath}	
\usepackage{amssymb}	
\usepackage{graphicx}   

\graphicspath{{figs/}}

\defcitealias{mason2018a}{M18a}  
\defcitealias{ren2019}{R19}      

\newcommand{\z}[1]{$z \sim #1$}  
\newcommand{\xHI}{$\overline{x}_{\ion{H}{i}}$}  
\newcommand{\Lya}{Ly$\alpha$}  
\newcommand{\Muv}{$M_\textsc{uv}$}  
\newcommand{\Mh}{$M_\text{h}$}  
\newcommand{\Tigm}{$\mathcal{T}_\textsc{igm}$}  
\newcommand{\xHIscatter}{$\overline{x}_{\ion{H}{i}} = 0.55_{-0.13}^{+0.11}$}  
\newcommand{\xHInoscatter}{$\overline{x}_{\ion{H}{i}} = 0.59_{-0.14}^{+0.12}$}  

\title[Ly$\alpha$ Visibility During the EoR]{The Impact of Scatter in the Galaxy UV Luminosity to Halo Mass Relation on Ly$\alpha$ Visibility During the Epoch of Reionization}

\author[Whitler et al.]{
Lily R. Whitler,$^{1}$\thanks{E-mail: \href{mailto:lwhitler@asu.edu}{lwhitler@asu.edu}}
Charlotte A. Mason,$^{2}$\thanks{Hubble Fellow}
Keven Ren,$^{3,4}$
Mark Dijkstra,
\newauthor
Andrei Mesinger,$^{5}$
Laura Pentericci,$^{6}$
Michele Trenti,$^{3,4}$
and Tommaso Treu$^{7}$
\\
$^{1}$School of Earth and Space Exploration, Arizona State University, 781 Terrace Mall, Tempe, AZ 85287, USA \\
$^{2}$Center for Astrophysics~\textbar~Harvard \& Smithsonian, 60 Garden St., Cambridge, MA 02138, USA \\
$^{3}$School of Physics, University of Melbourne, Parkville, Victoria, Australia \\
$^{4}$ARC Centre of Excellence for All Sky Astrophysics in 3 Dimensions (ASTRO 3D) \\
$^{5}$Scuola Normale Superiore, Piazza dei Cavalieri 7, I-56126 Pisa, Italy \\
$^{6}$INAF, Osservatorio Astronomico di Roma, via Frascati 33, I-00040 Monteporzio, Italy \\
$^{7}$Department of Physics and Astronomy, University of California, Los Angeles, CA 90095-1547, USA
}

\date{Accepted XXX. Received YYY; in original form ZZZ}

\pubyear{2020}

\begin{document}
\label{firstpage}
\pagerange{\pageref{firstpage}--\pageref{lastpage}}
\maketitle

\begin{abstract}
The reionization of hydrogen is closely linked to the first structures in the universe, so understanding the timeline of reionization promises to shed light on the nature of these early objects. In particular, transmission of Lyman alpha (\Lya) from galaxies through the intergalactic medium (IGM) is sensitive to neutral hydrogen in the IGM, so can be used to probe the reionization timeline. In this work, we implement an improved model of the galaxy UV luminosity to dark matter halo mass relation to infer the volume-averaged fraction of neutral hydrogen in the IGM from \Lya\ observations. Many models assume that UV-bright galaxies are hosted by massive dark matter haloes in overdense regions of the IGM, so reside in relatively large ionized regions. However, observations and N-body simulations indicate that scatter in the UV luminosity--halo mass relation is expected. Here, we model the scatter (though we assume the IGM topology is unaffected) and assess the impact on \Lya\ visibility during reionization. We show that UV luminosity--halo mass scatter reduces \Lya\ visibility compared to models without scatter, and that this is most significant for UV-bright galaxies. We then use our model with scatter to infer the neutral fraction, \xHI, at \z{7} using a sample of Lyman-break galaxies in legacy fields. We infer \xHIscatter\ with scatter, compared to \xHInoscatter\ without scatter, a very slight decrease and consistent within the uncertainties. Finally, we place our results in the context of other constraints on the reionization timeline and discuss implications for future high-redshift galaxy studies.
\end{abstract}

\begin{keywords}
dark ages, reionization, first stars -- galaxies: evolution -- galaxies: high-redshift -- intergalactic medium
\end{keywords}



\section{Introduction} \label{sec:intro}

The Epoch of Reionization (EoR) is currently one of the most poorly understood eras in cosmic history. During the EoR, neutral hydrogen in the intergalactic medium (IGM) was (re)ionized by the first sources of light in the universe -- stars and accreting black holes in galaxies. The properties of these sources are presently only loosely constrained, but a detailed understanding of the timeline of cosmic reionization will shed light on the characteristics of these early luminous objects \citep[e.g.][]{barkana2001, robertson2010, mesinger2016a, dayal2018, mason2019b}.

In the last decade, observations of the cosmic microwave background (CMB), gamma ray bursts, quasars, and galaxies have begun to probe this era. For example, \textit{Planck} measurements of the Thomson optical depth imply that hydrogen was 50 per cent neutral at $z \sim 7.8 - 8.8$ \citep{planck2016a, planck2016b}. Absorption features in quasar spectra, such as Gunn-Peterson troughs \citep{gunn1965} in $z \gtrsim 6$ spectra \citep{fan2006} and the dark pixel fraction in the Lyman alpha (\Lya; 1216\,\AA) and Lyman beta (Ly$\beta$) forest at $z \sim 5 - 6$ \citep{mcgreer2015}, suggest that reionization was largely complete by \z{6}, although there is growing evidence that reionization was incomplete as late as \z{5.5} \citep[e.g.][]{lidz2007, mesinger2010, kulkarni2019a}. Meanwhile, \Lya\ damping wing absorption in two quasar spectra at $z \gtrsim 7$ indicate that the universe was still partially neutral at \z{7} \citep[e.g.][]{greig2017, greig2019, banados2018, davies2018}.

However, probing the EoR with galaxies has advantages that CMB and quasar measurements lack. The CMB provides only integral constraints on the reionization history \citep{komatsu2011}, and bright quasars at $z \gtrsim 7$ are both extremely rare \citep[e.g.][]{parsa2018, kulkarni2019b} and potentially biased towards high-density regions \citep{lidz2007}. In contrast, galaxies can target specific stages of the EoR, and are more common and less biased than quasars at $z \gtrsim 7$.

In particular, \Lya\ emission from high-redshift galaxies is a key EoR probe. \Lya\ has an extremely large cross section for scattering by neutral hydrogen, making it sensitive to the ionization state of the IGM. Spectroscopic follow-up of galaxies selected as photometric dropouts (Lyman-break galaxies, or LBGs) show larger fractions of LBGs emitting \Lya\ with increasing redshift until \z{6} \citep{stark2010, hayes2011, curtis-lake2012, cassata2015}, likely due to decreasing dust fractions and thus less attenuation in the interstellar medium (ISM) \citep{hayes2011}. However, at $z \gtrsim 6$, the fraction of \Lya\ emitters begins to drop rapidly, suggesting increasing absorption of \Lya\ in the IGM as the universe becomes increasingly neutral \citep[e.g.][]{stark2010, pentericci2011, ono2012, schenker2012, treu2012}. Moreover, the clustering of \Lya-emitting galaxies also provides insights into the EoR. \Lya\ emission is preferentially transmitted through large ionized regions, boosting the clustering signal of \Lya\ emitters during reionization \citep{sobacchi2015, ouchi2018, weinberger2019}.

Quantitatively constraining the IGM ionization state with \Lya\ is nontrivial, however. The resonance of \Lya\ with neutral hydrogen makes it sensitive to galactic properties (e.g. neutral hydrogen column density and covering fraction) as well as the IGM \citep[e.g.][]{haiman1999, malhotra2004, santos2004, mcquinn2007, dijkstra2014, mason2018a}. Therefore, models must account for physics ranging from pc to Gpc scales.

In this work, we build on the methods of \citet{mason2018a} \citepalias[hereafter][]{mason2018a} to use \Lya\ observations to improve our understanding of the timeline of the EoR. \citetalias{mason2018a} model \Lya\ visibility during reionization by combining semi-numerical simulations of the IGM with empirical models of radiative transfer effects in the ISM. They obtain simulated cubes of the reionizing IGM and the distribution of dark matter haloes from the Evolution of 21\,cm Structure simulations \citep{mesinger2016b}, then use physically-motivated models to populate haloes with galaxy properties -- UV luminosities and \Lya\ line profiles -- to study transmission of \Lya\ through the reionizing IGM.

A key assumption of the \citetalias{mason2018a} models is a one-to-one mapping between UV luminosity and halo mass. \citetalias{mason2018a} use the model of \citet{mason2015}, which calibrates a star-formation efficiency at a single redshift with abundance matching, and evolves the UV luminosity--halo mass relation over cosmic time assuming star formation rates are proportional to halo mass accretion rates. This evolution preserves the one-to-one relation between UV luminosity and halo mass from the calibration redshift, and captures the simple physical intuition that UV luminous galaxies reside in high-mass haloes in more overdense regions, which reionize early.

To test this assumption, we now model \Lya\ visibility during the EoR with scatter in the galaxy UV luminosity--halo mass relation, which allows haloes of a single mass to host galaxies with a distribution of UV luminosities. Scatter is expected due to (at least) variation in halo mass assembly times and the stochastic nature of star formation, both of which increase with increasing redshift. \citet{ren2018} investigated the impact of scatter in the UV luminosity--halo mass relation in the context of the \citet{mason2015} modelling framework and calculated a minimum log-normal scatter between UV luminosity and halo mass at \z{7} of $\sim 0.5$\,mag. This is consistent with scatter obtained in semi-analytical models built on N-body simulations \citep[e.g.][]{behroozi2010, behroozi2013, moster2013, ren2018, tacchella2018}, estimated from comparing observed luminosity and stellar mass functions \citep{finkelstein2015}, and satellite kinematics \citep{more2009}. This scatter is also consistent with the observed $z \gtrsim 6$ luminosity functions \citep{ren2019}.

Scatter in the UV luminosity--halo mass relation alters the distribution of galaxies within the reionizing IGM, and therefore impacts \Lya\ transmission through the IGM. As the most massive haloes are extremely rare, including scatter greatly increases the probability of a very luminous galaxy being hosted in a more common, lower-mass halo. In contrast, abundance matching places UV-bright galaxies in high-mass haloes, which tend to reside in overdense regions with abundant neighboring haloes, creating larger ionized bubbles \citep[e.g.][]{furlanetto2004, furlanetto2006, mcquinn2007} and facilitating higher transmission of \Lya\ through the IGM. But if the most UV-luminous galaxies do not necessarily live in the most massive haloes surrounded by large ionized regions, and rather can reside in less massive haloes surrounded by smaller ionized bubbles, this could reduce the visibility of \Lya\ from these galaxies. On average, if the degree of absorption in the IGM is fixed, this would tend to reduce the quantity of neutral hydrogen in the IGM necessary to attenuate \Lya\ to the same extent. That is, the UV luminosity--halo mass relation without scatter requires a higher volume-averaged fraction of neutral hydrogen in the IGM to suppress \Lya\ than when scatter is included.

This paper is structured as follows. We describe our model for \Lya\ visibility during the EoR, including scatter in the UV luminosity--halo mass relation, in Section~\ref{sec:modelling}. In Section~\ref{sec:results}, we present our results on \Lya\ visibility and our inferred value of the fraction of neutral hydrogen in the IGM at \z{7}, accounting for scatter. We discuss our results in Section~\ref{sec:discussion}, and summarise and conclude in Section~\ref{sec:conclusion}.

We adopt a $\Lambda$CDM cosmology with parameters $\Omega_\Lambda = 0.69$, $\Omega_\text{m} = 0.31$, $\Omega_\text{b} = 0.048$, $H_0 = 68$\,km\,s$^{-1}$\,Mpc$^{-1}$, $n_\text{s} = 0.97$, and $\sigma_8 = 0.81$ \citep{planck2016a}. We use the \citet{sheth2001} halo mass function and all magnitudes are given in the AB system.

\section{Modelling \texorpdfstring{L\MakeLowercase{y}$\alpha$}{Lya} Visibility} \label{sec:modelling}

To model the visibility of \Lya\ during the EoR, we combine simulations of the IGM from the Evolution of 21\,cm Structure\footnote{\url{http://homepage.sns.it/mesinger/EOS.html}} (EOS) project \citep{mesinger2016b} with empirical models of ISM properties. We have revised the treatment of the galaxy UV luminosity to halo mass relation from the methods described by \citetalias{mason2018a}, though we refer the reader there for most details. Below, we provide a brief overview of the full model and a full description of our implementation of the galaxy UV luminosity--halo mass relation with scatter, which follows the conditional luminosity function modelling of \citet{ren2019} (hereafter \citetalias{ren2019}).

\subsection{\texorpdfstring{Ly$\alpha$}{Lya} Transmission Through the ISM and IGM} \label{subsec:ISM_IGM_modelling}

We use the same simulations of the reionizing IGM and dark matter haloes as those used by \citetalias{mason2018a}, obtained from the EOS simulations \citep{mesinger2016b}. The EOS simulations use \textsc{21cmfastv2} \citep{sobacchi2014}, which incorporates sub-grid prescriptions for inhomogeneous recombinations and photoheating feedback, to simulate cubes (1.6 comoving\,Gpc on a side with resolution of $1024^3$) of the IGM ionization state and 21\,cm signal during Cosmic Dawn and the EoR \citep{mesinger2016b}. A specific reionization history is not assumed; rather, the ionization field at a given redshift is generated directly from a nonlinear density field assuming an ionizing efficiency parameter. Additionally, we note that we do not expect recombinations to considerably impact the morphology of reionization, since they only contribute significantly in the final $\lesssim 10$ per cent of reionization \citep{sobacchi2015}.

The ionizing efficiency is proportional to the product of the escape fraction of ionizing photons and the stellar mass fraction in a halo. Since the escape fraction likely decreases with increasing halo mass while the stellar mass fraction decreases, the ionizing efficiency is approximately constant when averaged over all haloes. Then, the ionizing efficiency is chosen to ensure that the simulated Thompson scattering optical depth is consistent with that measured by \citet{planck2016a}. See \citet{mesinger2007, mesinger2011} and \citet{mesinger2015} for further details.

To obtain various values of \xHI\ at a \z{7}, we apply ionization fields from different redshifts to the density field at \z{7}, analogous to varying the escape fraction of ionizing photons and the timing of reionization \citep[e.g.][]{mcquinn2007, jensen2014, sobacchi2016}. While this is not fully self-consistent, the topology of reionization is highly redshift-independent and instead depends primarily on the global neutral fraction \citep{mcquinn2007}.

The sources of reionization are taken to be star-forming galaxies hosted by dark matter haloes of some spatially-dependent minimum mass, $M_\text{min}(\boldsymbol{x}, z)$. The minimum mass is determined by taking the maximum of the halo mass corresponding to the virial temperature required for cooling by atomic hydrogen ($M_\text{cool}$, assuming a virial temperature of $T_\text{vir} \gtrsim 2 \times 10^4$\,K), the halo mass required to form stars inside \ion{H}{ii} regions ($M_\text{photo}$), and the halo mass determined by supernova feedback ($M_\text{SNe}$), since more efficient feedback suppresses star formation for higher-mass haloes \citep{mesinger2016b}.

The role of supernova feedback in star formation during reionization is largely unconstrained, so \citet{mesinger2016b} take $M_\text{SNe}$ to be a free parameter and choose two extreme values to approximately encompass the parameter space of the sources that drive reionization. In the `Faint Galaxies' run, supernova feedback is inefficient, $M_\text{min}(\boldsymbol{x}, z)$ is fixed by $M_\text{cool}$ and $M_\text{photo}$, and the primary drivers of reionization are low-mass galaxies of halo mass $\sim 10^{8} - 10^{9}$\,M$_{\sun}$ \citep[see Figure 1 of][]{mesinger2016b} that produce small ionized regions. In contrast, the `Bright Galaxies' run corresponds to extremely efficient supernova feedback and is characterised by more massive galaxies (halo mass $\sim10^{10}$\,M$_{\sun}$) that produce larger ionized patches. In this work, we choose the `Faint Galaxies' run as our fiducial model, as the difference in the inferred volume-averaged neutral fraction between the two is negligible for current observational samples \citep{mason2018a, mason2018b, greig2019}.

We populate the simulated dark matter haloes (with mass $10^{10} \leq M_\text{h}/\text{M}_{\sun} \leq 10^{12}$) with UV magnitudes (\Muv), and \Lya\ line profiles and emitted \Lya\ equivalent widths (EWs) drawn from empirical models, and calculate the observed \Lya\ EW after transmission along sightlines through the reionizing IGM. We describe our method for assigning UV magnitudes, updated from the models of \citetalias{mason2018a}, in Section~\ref{subsec:CLF}. Otherwise, we follow the procedure presented by \citetalias{mason2018a} in section~2.1 (and which we briefly outline here) to supply the haloes with emitted \Lya\ EWs in the rest frame and \Lya\ line profiles.

To model the observed \Lya\ EW distribution at \z{7}, we first assume that the observed \z{6} EW distribution is equal to the emitted \z{7} EW distribution. That is, any change in the observed \Lya\ EW distribution between \z{7} and \z{6} is entirely due to reionization and not evolution of galactic properties such as dust. This is likely a simplification, as lower-redshift galaxies show a trend towards increasing \Lya\ EWs with increasing redshift as dust fraction decreases \citep{hayes2011}. However, the time-scale between \z{7} and \z{6} is short ($\lesssim 200$ Myr) and we do not expect significant change \citep{treu2012}. Moreover, if the intrinsic emitted \Lya\ EW distribution does evolve significantly between \z{7} and \z{6}, it will likely be towards higher EWs at \z{7} due to decreasing dust \citep{hayes2011}, requiring a higher neutral fraction than we find to suppress the modelled \Lya\ EW distribution to that which is observed. Given this caveat (discussed further in Section~\ref{subsec:modelling_caveats}), we proceed with the model for the emitted \Lya\ EW distribution at \z{7}, $p(\text{EW}_\text{emit} \mid M_\textsc{uv})$, derived by \citetalias{mason2018a} from a fit to the observed \z{6} sample by \citet{debarros2017} (the largest homogeneous sample of \z{6} galaxies with deep spectroscopic observations probing the presence of \Lya\ for EW$_{\text{Ly}\alpha} \geq 25$\ \AA).

We use the differential \Lya\ transmission fraction, \Tigm, to compare the \Lya\ EW distributions at \z{7} and \z{6}. \Tigm\ is defined as the fraction of emitted \Lya\ photons that are transmitted through the IGM, $\mathcal{T}_\textsc{igm} = \text{EW}_\text{obs} / \text{EW}_\text{emit}$, and can be calculated as
\begin{equation} \label{eqn:Tigm}
\mathcal{T}_\textsc{igm}(\overline{x}_{\ion{H}{i}}, M_\text{h}, \Delta v) =
\int_{-\infty}^\infty \text{d}v \, J_{\text{Ly}\alpha}(\Delta v, M_\text{h}, v) \, \text{e}^{-\tau_\textsc{igm}(\overline{x}_{\ion{H}{i}}, M_\text{h}, v)},
\end{equation}
where \Tigm\ is normalised to $\mathcal{T}_{\textsc{igm}}(z = 6) = 1$. $J_{\text{Ly}\alpha}$ is the line profile of \Lya\ photons escaping from galaxies as a function of velocity, $v$, including the velocity offset of the \Lya\ line from the systemic velocity of the galaxy ($\Delta v$, see below), and $\tau_\textsc{igm}$ is the optical depth to \Lya\ photons through the IGM. $\tau_\textsc{igm}$ is comprised of two components: the \Lya\ damping wing absorption in the partially neutral IGM, $\tau_\textsc{d}(\overline{x}_{\ion{H}{i}}, M_\text{h}, v)$, and resonant absorption due to residual neutral gas in the local ionized region around the source, $\tau_{\ion{H}{ii}}(M_\text{h}, v)$ \citep[see e.g.][for a review]{dijkstra2014}. While $\tau_\textsc{d}$ only arises during reionization, $\tau_{\ion{H}{ii}}$ can be significant at all redshifts, and increases with redshift due the increasing density of gas \citep{laursen2011}. In addition, infall of over-dense gas around haloes causes significant resonant absorption of \Lya\ photons on the red side of line center \citep{santos2004,dijkstra2007,laursen2011,weinberger2018}. In this work, following \citet{santos2004} and \citetalias{mason2018a}, we model $\tau_{\ion{H}{ii}}$ as a step function, truncated below the halo's circular velocity to model resonant absorption in the infalling circumgalactic medium (CGM) and the ionized IGM.

\citet{dijkstra2011} first showed that radiative transfer effects within galaxies, particularly scattering in galactic outflows, strongly impacts the emerging \Lya\ line profile, which in turn affects the transmission of \Lya\ through the reionizing IGM. In this work, we model radiative transfer in the ISM using the methods of \citetalias{mason2018a}, which models the line shape of \Lya\ emerging from the ISM as a Gaussian centred at a velocity offset, $\Delta v$, from the systemic redshift of the source. In our empirical model, $\Delta v$ and halo mass are correlated, as lower-mass galaxies are expected to have a lower column density of neutral hydrogen than high-mass galaxies \citep[e.g.][]{yang2017}, so scattering in the ISM is decreased \citep[e.g.,][]{neufeld1990}. Additionally, $\Delta v$ is often correlated with the outflow velocity \citep{verhamme2006, verhamme2008}, and thus the halo's circular velocity, which is also regulated by the halo mass. We note that in this formulation, we have assumed that the evolution of \Tigm\ is only due to the changing \Lya\ damping wing optical depth. That is, we do not model any redshift evolution (at fixed \Mh) of absorption by residual neutral hydrogen in the ionized component of the IGM and CGM, or due to evolving $\Delta v$. See Section \ref{subsec:modelling_caveats} for further discussion.

We compute \Lya\ optical depths, $\tau_\textsc{igm}$, for $\sim 10^4$ sightlines through the simulated IGM for a range of neutral fractions from $0.01 \leq \overline{x}_{\ion{H}{i}} \leq 0.95$ with $\Delta \overline{x}_{\ion{H}{i}} \approx 0.02$ (see \citetalias{mason2018a} for details) and evaluate \Tigm\ with Equation \ref{eqn:Tigm}. We then calculate the observed \Lya\ EW, $\text{EW}_\text{obs}$, as $\text{EW}_\text{obs} = \mathcal{T}_\textsc{igm} \times \text{EW}_\text{emit}$, where $\text{EW}_\text{emit}$ is drawn from the \Muv-dependent emitted \Lya\ EW distribution for $10^5$ simulated galaxies.

Using our sample of \Tigm\ values, we calculate probability density functions for \Tigm\ given \Muv, \xHI, and scatter in the UV luminosity--halo mass relation ($\sigma$) (discussed further in Section~\ref{subsec:CLF}): $p(\mathcal{T}_\textsc{igm} \mid M_\textsc{uv}, \overline{x}_{\ion{H}{i}}, \sigma)$. We calculate $p(\mathcal{T}_\textsc{igm} \mid M_\textsc{uv}, \overline{x}_{\ion{H}{i}}, \sigma)$ on a grid of neutral fractions and UV magnitudes ($-21.9 \leq M_\textsc{uv} \leq -16$, $\Delta M_\textsc{uv} = 0.1$). To do this, we first evaluate $p(\mathcal{T}_\textsc{igm} \mid M_\text{h}, \overline{x}_{\ion{H}{i}})$ on a grid of neutral fractions and halo masses ($10.2 \lesssim \log M_\text{h} \lesssim 11.9$, $\Delta \log M_\text{h} \approx 0.1$), and map from halo mass to UV magnitude using the method described below.

\begin{figure*}
    \centering
    \includegraphics[width=\textwidth]{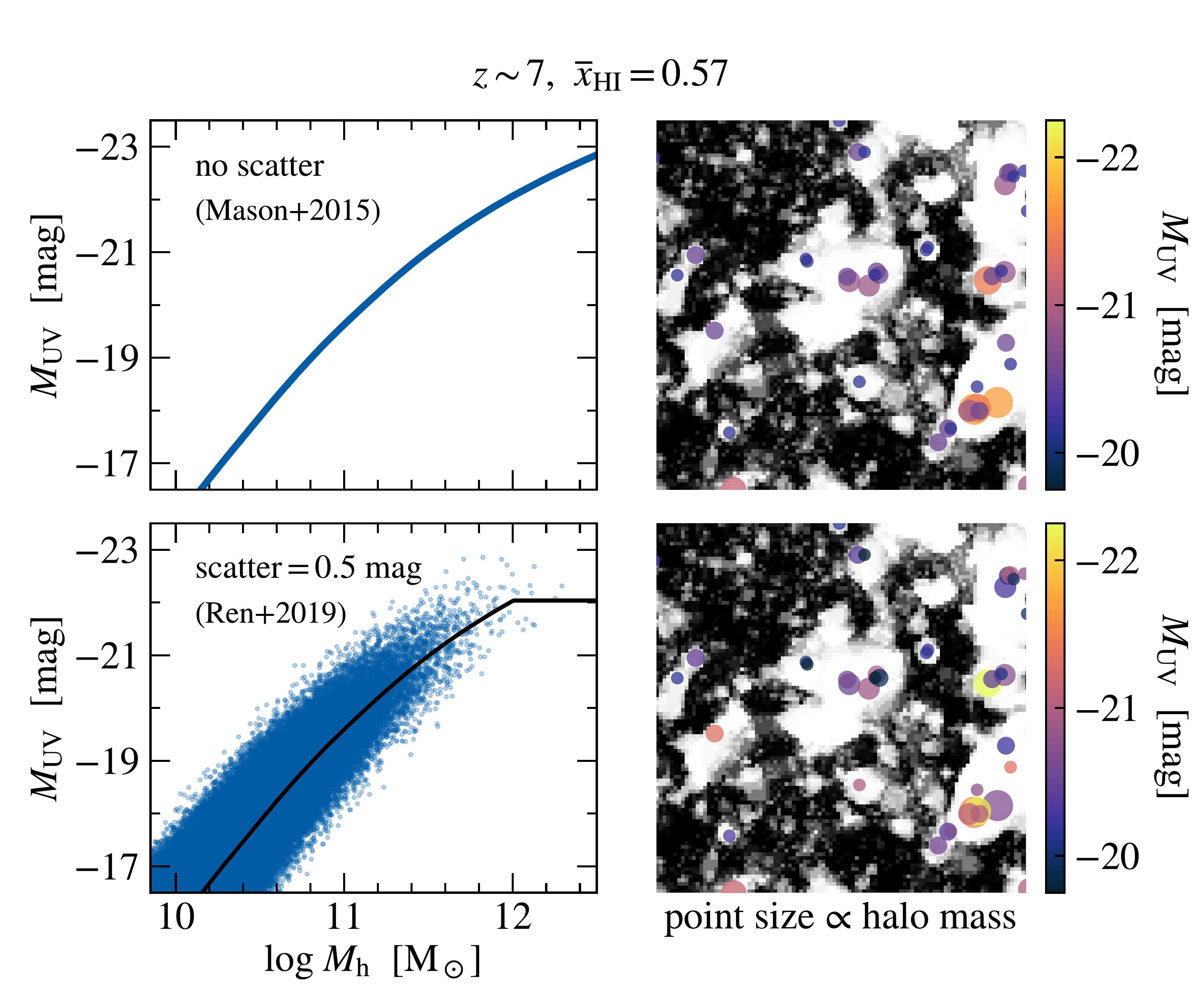}
	\caption{The spatial distribution of dark matter haloes in a $150 \times 150 \times 5$\,Mpc slice of the reionizing IGM at \z{7} (with volume-averaged fraction of neutral hydrogen of $\overline{x}_{\ion{H}{i}} = 0.57$), and the UV magnitudes of their corresponding galaxies with and without scatter in the UV magnitude to halo mass relation. Left: The UV magnitude--halo mass relations at \z{7}. The relation without scatter (top) is derived from the technique based on abundance matching described by \citet{mason2015}, and the relation with scatter of $\sigma = 0.5$\,mag (bottom) follows the conditional luminosity function presented by \citet{ren2019}, where the solid black line is the median UV magnitude--halo mass relation. Right: The spatial distribution of haloes (point size is proportional to halo mass) in the reionizing IGM with $\overline{x} = 0.57$ (white is ionized and black is neutral), simulated with \textsc{21cmfast} \citep{mesinger2007, mesinger2011}. For clarity, only haloes of mass $M_\text{h} \geq 10^{11}$\,M$_{\sun}$ are shown. The galaxy UV magnitudes, assigned with the UV magnitude--halo mass relation in the corresponding left panel, are shown as the colour scale. Without scatter, the most massive haloes in the largest ionized bubbles host the most UV-bright galaxies. In contrast, lower-mass haloes in smaller ionized regions can host more UV-luminous galaxies when scatter is included.}
	\label{fig:scatter_xHI_map}
\end{figure*}

\subsection{Incorporating Scatter in the UV Luminosity--Halo Mass Relation} \label{subsec:CLF}

To model \Lya\ emission from galaxies, we map from observed UV luminosity to emitted \Lya\ EW and differential \Lya\ transmission fraction, based on dark matter halo mass. This requires modelling the relationship between halo mass and UV luminosity. \citetalias{mason2018a} use the one-to-one model described by \citet{mason2015} (which successfully reproduces $0 \lesssim z \lesssim 10$ UV luminosity functions assuming star formation rates are proportional to halo mass accretion rates modulo a mass-dependent but redshift-independent efficiency) to map from halo mass to UV luminosity. We now include scatter in the galaxy UV luminosity--halo mass relation using the conditional luminosity function (CLF) approach of \citetalias{ren2019}, which introduces scatter in \Muv\ given a halo mass, \Mh.

The CLF presented by \citetalias{ren2019} is a log-normal distribution of galaxy UV luminosities, $L$, given a host halo mass, \Mh, with a median UV luminosity--halo mass relation, $L_c$, and a log-normal dispersion, $\sigma_{\log L}$. We have converted the CLF of \citetalias{ren2019} from log-normal in luminosity to normal in magnitude, so $\log L$ becomes \Muv, $\log L_c$ becomes $M_{\textsc{uv}, c}$, and the standard deviation is $\sigma = \sigma_{M_\textsc{uv}} = 2.5\,\sigma_{\log L}$. Thus, our CLF is as follows:
\begin{equation} \label{eqn:Muv_Mh_scatter}
p(M_\textsc{uv} \mid M_\text{h}) = \frac{1}{\sqrt{2\pi}\sigma} \exp \left(\frac{-[M_\textsc{uv} - M_{\textsc{uv}, c}(M_\text{h}, \sigma, z)]^2}{2\sigma^2}\right).
\end{equation}
The dispersion, $\sigma$, was originally introduced to explain scatter observed in the Tully--Fisher relation \citep{yang2005}, and is a free parameter in the CLF.

The standard UV luminosity function, $\Phi(M_\textsc{uv})$, is derived from the CLF by convolving with the halo mass function:
\begin{equation}
\label{eqn:uvlf_clf}
\Phi(M_\textsc{uv}) = \int_0^\infty \text{d}M_\text{h} \, p(M_\textsc{uv} \mid M_\text{h}) \, \frac{\text{d}n}{\text{d}M_\text{h}},
\end{equation}
where $p(M_\textsc{uv} \mid M_\text{h})$ is given by Equation \ref{eqn:Muv_Mh_scatter} and $\text{d}n/\text{d}M_\text{h}$ is the halo mass function (we use the \cite{sheth2001} halo mass function). The luminosity function is calibrated to the observed UV luminosity function at \z{5} \citep{bouwens2015} and extrapolated based on the galaxy evolution model of \citet{mason2015}. For $\sigma > 0$, the median UV luminosity to halo mass relation is then a function of halo mass, standard deviation ($\sigma$), and redshift, and a critical flattening threshold is imposed based on either mass or luminosity. The flattening threshold can be linked to feedback processes by active galactic nuclei; an extended discussion is given in \citetalias{ren2019}.

The degree of scatter, $\sigma$, is a free parameter in the CLF model. However, \citetalias{ren2019} explored a range of values for $\sigma$ and found that $\sigma = 0.5$\,mag is consistent with the observed luminosity functions for $z = 6 - 10$ in the framework of the \citet{mason2015} UV luminosity function evolution model. Figure 4 of \citetalias{ren2019} shows a comparison of the modelled luminosity functions with scatter of $\sigma = 0, 0.5, 1$\,mag for selected redshifts from $z \sim 6 - 12$, demonstrating that both no scatter and scatter of $\sigma = 0.5$\,mag are consistent with observations, while $\sigma = 1$\,mag overpredicts the bright end. Thus, we use $\sigma = 0.5$\,mag as our fiducial model to ensure that we reproduce the observed UV luminosity function at \z{7}.

The introduction of scatter into the \citet{mason2015} evolution model with the CLF requires a critical flattening threshold to maintain consistency with observations. This flattening can be regulated by either a critical mass threshold or a critical luminosity threshold, and the predicted luminosity functions are consistent with the observed luminosity functions for $z = 6 - 10$ for both (see figure 4 of \citetalias{ren2019}). We tested the impact on the inference in Section~\ref{subsec:inferred_xHI} of imposing a critical luminosity flattening threshold instead of a critical mass threshold, and found the difference in the inferred neutral fraction was negligible. This is expected, as at \z{7}, there is little difference in the median UV luminosity--halo mass relation between the two flattening criteria \citepalias{ren2019}. In this work, we choose to use the critical mass flattening threshold, though either criteria would suffice since the modelled UV luminosity function and our inferred neutral fraction is insensitive to the choice of flattening threshold.

To implement the scatter, we use the \textsc{hmf}\footnote{\url{https://github.com/steven-murray/hmf}} package \citep{murray2013} to sample millions of haloes ranging from $\sim 10^{10} - 10^{12.5}$\,M$_{\sun}$ from the \citet{sheth2001} halo mass function at \z{7}, and assign each halo a UV magnitude according to the CLF \citepalias[Equation \ref{eqn:Muv_Mh_scatter};][]{ren2019}. Fig.\ \ref{fig:scatter_xHI_map} shows the spatial distribution of haloes in a $150 \times 150 \times 5$\,Mpc slice of the partially reionized IGM at \z{7}, simulated with \textsc{21cmfast}\footnote{\url{https://github.com/andreimesinger/21cmFAST}} \citep{mesinger2007, mesinger2011}, where the ionizing efficiency is set to obtain an average neutral fraction of $\overline{x}_{\ion{H}{i}} \approx 0.60$ at \z{7}. We note this is only an illustration of our model, as the EOS simulations we use for this work are larger (1.6\,comoving\,Gpc on a side) and include prescriptions for inhomogeneous recombinations and photoheating feedback; see section~2.1 of \citet{mesinger2016b}. We also show the UV magnitudes of the galaxies hosted by the haloes, assigned with and without scatter, as the colour scale. Without scatter, UV-bright galaxies reside in high-mass haloes surrounded by large bubbles of ionized hydrogen. Conversely, with scatter, UV-bright galaxies can be hosted in lower-mass haloes surrounded by smaller ionized regions.

\subsubsection{\texorpdfstring{Ly$\alpha$}{Lya} Transmission Probability} \label{subsubsec:Mh_marginalization}

As \Muv\ is a real observable, we wish to obtain $p(\mathcal{T}_\textsc{igm} \mid M_\textsc{uv}, \overline{x}_{\ion{H}{i}}, \sigma)$ (where we have chosen $\sigma = 0.5$\,mag), but the outputs of our model are $p(\mathcal{T}_\textsc{igm} \mid M_\text{h}, \overline{x}_{\ion{H}{i}})$ (as described in Section~\ref{subsec:ISM_IGM_modelling}). Accordingly, we marginalise over halo mass to calculate probability distributions of the differential \Lya\ transmission fraction, \Tigm, for fixed UV magnitude, \Muv, neutral fraction, \xHI, and scatter, $\sigma$:
\begin{align} \label{eqn:marginal_pTigmMh}
p(\mathcal{T}_\textsc{igm} \mid M_\textsc{uv}, \overline{x}_{\ion{H}{i}}, \sigma) = & \\ \int_0^\infty & \text{d}M_\text{h} \, p(\mathcal{T}_\textsc{igm} \mid M_\text{h}, \overline{x}_{\ion{H}{i}}) \, p(M_\text{h} \mid M_\textsc{uv}, \sigma), \nonumber
\end{align}
where $p(\mathcal{T}_\textsc{igm} \mid M_\text{h}, \overline{x}_{\ion{H}{i}})$ is the probability distribution of \Tigm\ given a halo mass, \Mh, and neutral fraction, \xHI, as generated by \citetalias{mason2018a}, and we use Equation \ref{eqn:Muv_Mh_scatter} to find $p(M_\text{h} \mid M_\textsc{uv}, \sigma)$.

We use a numerical approach to calculate $p(M_\text{h} \mid M_\textsc{uv}, \sigma)$ from Equation \ref{eqn:Muv_Mh_scatter} for $\sigma = 0.5$\,mag. We create a two-dimensional histogram of $p(M_\textsc{uv} \mid M_\text{h})$ with bin sizes of $\Delta \log M_\text{h} \approx 0.1$ to match the EOS simulation halo masses, and $\Delta M_\textsc{uv} = 0.1$ to match our chosen \Muv\ grid. To obtain $p(M_\text{h} \mid M_\textsc{uv}, \sigma)$, we fix \Muv\ and extract the corresponding one-dimensional histogram.

\begin{figure}
    \centering
	\includegraphics[width=\columnwidth]{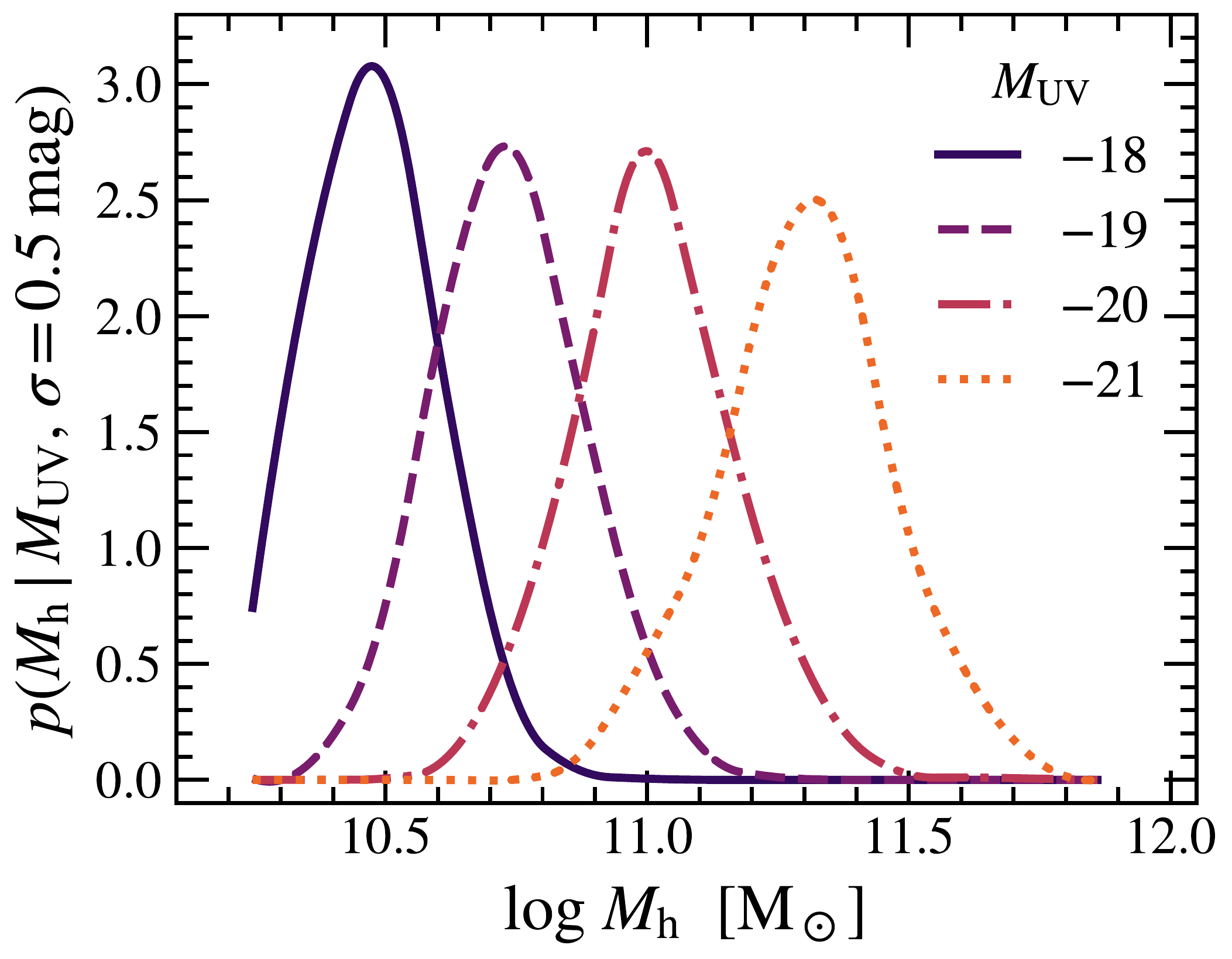}
	\caption{Normalised probability density functions of halo mass, \Mh, given UV magnitude, \Muv, and scatter of $\sigma = 0.5$\,mag (note that these distributions do not show relative number densities of galaxies of various UV magnitudes, which follow the UV luminosity function). Colours and linestyles correspond to galaxies of UV magnitudes $M_\textsc{uv} = -18, -19, -20, -21$, where darker colours indicate fainter galaxies. Though brighter galaxies are more likely than fainter galaxies to have high-mass hosts, there is significant scatter around their peak probability. There is also some broadening in the distributions, where bright galaxies may reside in a wider range of halo masses than faint ones.}
	\label{fig:pMh_Muv_hists}
\end{figure}

Sample distributions of $p(M_\text{h} \mid M_\textsc{uv}, \sigma)$ for a range of UV magnitudes are shown in Fig.\ \ref{fig:pMh_Muv_hists} (note that these distributions do not show relative number densities of different UV magnitudes). Bright galaxies are more likely than faint galaxies to reside in high-mass haloes, but with significant dispersion around the halo mass with peak probability. Additionally, there is slight broadening in the probability distributions from faint to bright galaxies, where bright galaxies may be hosted in a wider range of halo masses than fainter galaxies.

It should be noted that we have assumed that when scatter in the UV luminosity--halo mass relation is included, the distribution of galaxies within the IGM changes, but the IGM topology does not. This is reasonable if low-mass, faint galaxies dominate reionization, but if bright galaxies contribute significantly, the topology may change. Namely, bright galaxies in low-mass haloes may reside in larger ionized regions than we model. Thus, our results demonstrate the maximum expected deviation from the case without scatter. We discuss this assumption in more detail in Section~\ref{subsec:modelling_caveats}.

\section{Results} \label{sec:results}

In this section, we describe the key effects on \Lya\ visibility of modelling scatter in the UV luminosity to halo mass relation. In Section~\ref{subsec:lya_results}, we compare our results with scatter with the findings of \citetalias{mason2018a} without scatter in the UV luminosity to halo mass relation. In Section~\ref{subsec:inferred_xHI}, we infer the neutral fraction with observations presented by \citet{pentericci2014} and compare with the neutral fraction inferred by \citetalias{mason2018a}.

\subsection{\texorpdfstring{Ly$\alpha$}{Lya} Visibility} \label{subsec:lya_results}

\begin{figure*}
    \centering
	\includegraphics[width=\textwidth]{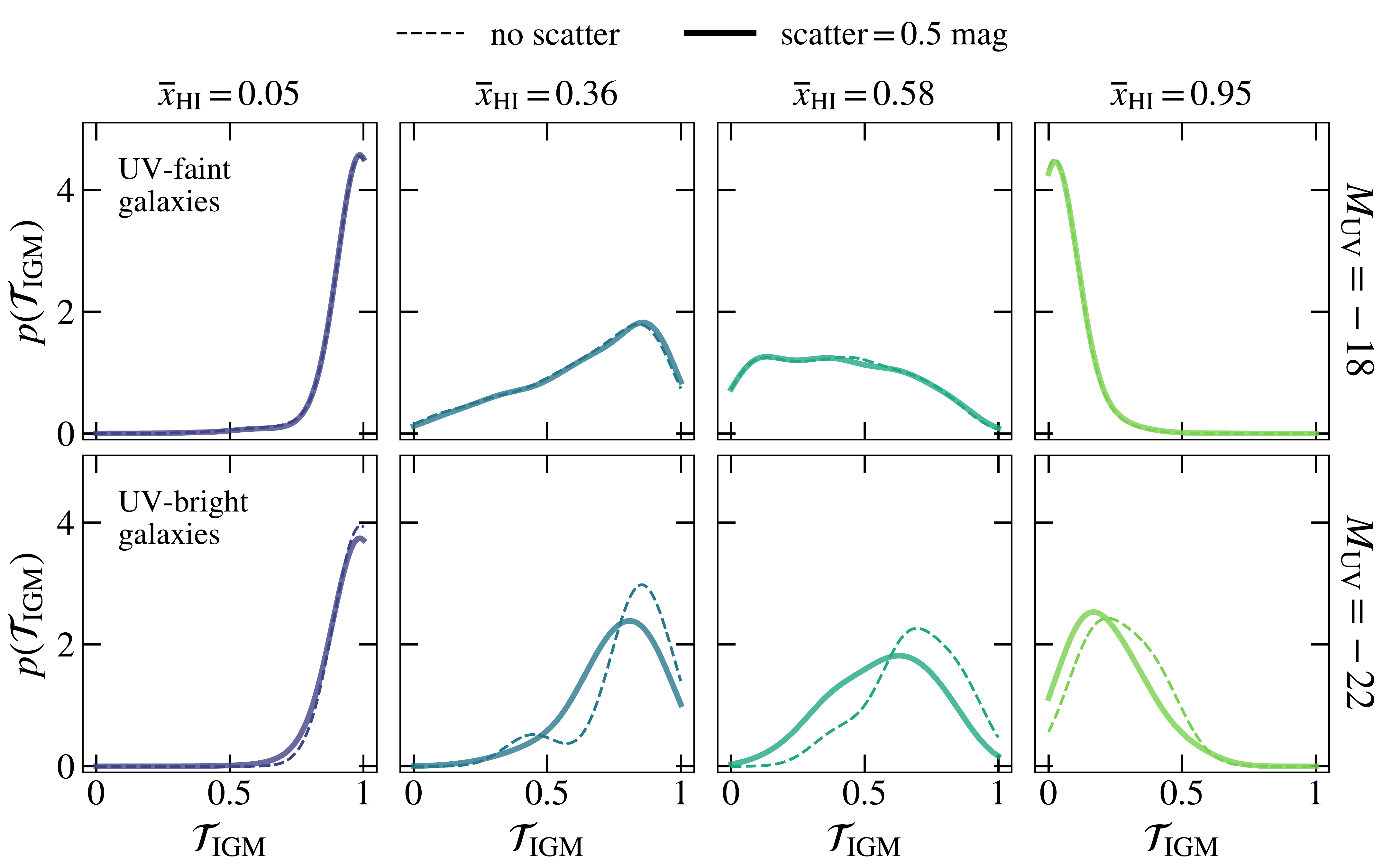}
	\caption{Comparison of the probability distributions of differential \Lya\ transmission fractions through the IGM, \Tigm, with scatter (solid) and without scatter (dashed) in the UV luminosity to halo mass relation at \z{7}. Each column and line colour shows a different neutral fraction, where darker colours correspond to lower neutral fractions (i.e. a more ionized IGM). UV-faint ($M_\textsc{uv} = -18$) galaxies are in the top row and UV-bright ($M_\textsc{uv} = -22$) galaxies are in the bottom row. Introducing scatter tends to increase the probability of lower transmission, particularly for UV-bright galaxies.}
	\label{fig:lya_transmission}
\end{figure*}

In Fig.\ \ref{fig:lya_transmission}, we plot probability density functions for \Tigm\ for two values of \Muv, mapping from \Mh\ to \Muv\ with and without incorporating scatter in the UV luminosity--halo mass relation. To transform from \Mh\ to \Muv\ without scatter, we use the one-to-one models of \citetalias{mason2018a}, which in turn use the UV luminosity function model of \citet{mason2015}. To transform from \Mh\ to \Muv\ with scatter, we use the CLF approach of \citetalias{ren2019} described in Section~\ref{subsec:CLF}, where we have chosen a dispersion of $\sigma = 0.5$\,mag. With scatter, \Lya\ transmission tends to decrease for UV-bright galaxies compared to the case without scatter.

\begin{figure}
    \centering
	\includegraphics[width=\columnwidth]{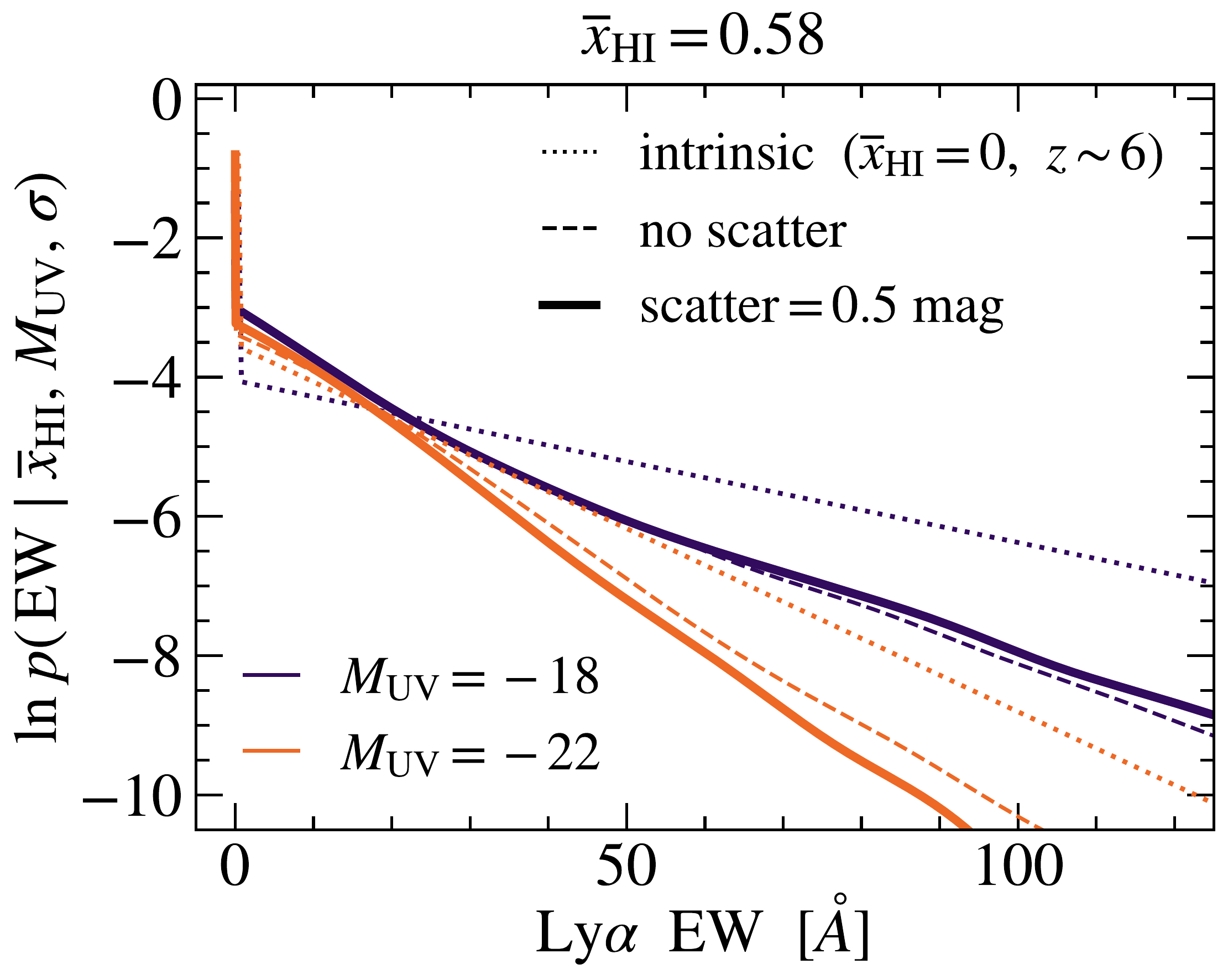}
	\caption{Probability distributions of \Lya\ EWs with (solid) and without (dashed) scatter of $\sigma = 0.5$\,mag in the UV luminosity--halo mass relation for UV-faint and UV-bright galaxies at one neutral fraction, $\overline{x}_{\ion{H}{i}} = 0.58$. UV-faint ($M_\textsc{uv} = -18$) galaxies are shown in purple and UV-bright ($M_\textsc{uv} = -22$) galaxies are shown in orange. The intrinsic distributions in a fully ionized universe ($\overline{x}_{\ion{H}{i}} = 0$) at \z{6} are shown as dotted lines. With scatter, bright galaxies tend to slightly lower \Lya\ EWs than without scatter, reflecting the decrease in \Lya\ transmission with scatter (Fig.\ \ref{fig:lya_transmission}).}
	\label{fig:lya_pW}
\end{figure}

Fig.\ \ref{fig:lya_pW} shows the probability distributions of observed \Lya\ EWs with and without scatter for faint and bright galaxies at one neutral fraction, $\overline{x}_{\ion{H}{i}} = 0.58$, calculated as described in Section~\ref{subsec:ISM_IGM_modelling}. The \Lya\ EW distribution for UV-bright galaxies tends towards slightly lower EWs, reflecting the decreased \Lya\ transmission for bright galaxies seen in Fig.\ \ref{fig:lya_transmission}.

For a fixed neutral fraction and UV luminosity, high \Lya\ transmissions and EWs are less probable when scatter in the UV luminosity--halo mass relation is included, especially for more UV-luminous objects. In general, modelling scatter allows UV-bright objects to reside in less massive haloes, and so less overdense regions. Hence, on average, they are surrounded by smaller ionized regions than would be predicted by an abundance matching technique. This tends to increase the attenuation of \Lya\ by neutral hydrogen along the line of sight.

\subsection{Inferred Neutral Fraction at \texorpdfstring{$z \sim 7$}{z7}} \label{subsec:inferred_xHI}

We infer the neutral fraction, \xHI, at \z{7} with the flexible Bayesian framework described by \citetalias{mason2018a}. We provide a brief description below, but refer the reader to \citetalias{mason2018a} for further details.

By Bayes' Theorem, we can write the posterior probability of \xHI\ inferred from one observation of a galaxy with measured \Lya\ EW (or upper limit), $\text{EW}_i$, and UV magnitude, $M_{\textsc{uv}, i}$, given the UV magnitude--halo mass scatter, $\sigma$, as the following:
\begin{equation}
p(\overline{x}_{\ion{H}{i}} \mid \text{EW}_i, M_{\textsc{uv}, i}, \sigma) \propto p(\text{EW}_i \mid M_{\textsc{uv}, i}, \overline{x}_{\ion{H}{i}}, \sigma) p(\overline{x}_{\ion{H}{i}}),
\label{eqn:xHI_posterior}
\end{equation}
where $p(\text{EW}_i \mid M_{\textsc{uv}, i}, \overline{x}_{\ion{H}{i}}, \sigma)$ is the \textit{likelihood} of observing a \Lya\ EW (or upper limit) given our model of \Lya\ visibility (including our chosen scatter of $\sigma = 0.5$\,mag in the UV luminosity to halo mass relation) and $p(\overline{x}_{\ion{H}{i}})$ is our \textit{prior} on \xHI, which we take to be uniform between 0 and 1.

We use a non-analytic prescription to obtain the likelihood of observing our data (sets of $\{\text{EW}, M_\textsc{uv}\}$ for galaxies at \z{7}) given our model of \Lya\ transmission through the ISM and IGM, as the complex topology of the IGM prevents an analytic formulation. We use our models, described in Section~\ref{subsec:CLF}, to generate large samples of mock observed \Lya\ EWs on a grid of neutral fractions, \xHI, and UV magnitudes, \Muv, and use a Gaussian Kernel Density Estimator \citep{rosenblatt1956, parzen1962} to generate smooth estimates for $p(\text{EW} \mid M_\textsc{uv}, \overline{x}_{\ion{H}{i}}, \sigma)$. We also account for uncertainties and upper limits of \Lya\ EW measurements as described by \citetalias{mason2018a}. Finally, we combine the inference from a set of $N_\text{gals}$ uncorrelated observations by multiplying the individual posterior distributions:

\begin{equation}
p(\overline{x}_{\ion{H}{i}} \mid \{\text{EW}, M
_\textsc{uv}\}, \sigma) \propto
\prod_{i = 1}^{N_\text{gals}} p(\text{EW}_i \mid M_{\textsc{uv}, i}, \overline{x}_{\ion{H}{i}}, \sigma) p(\overline{x}_{\ion{H}{i}}).
\label{eqn:full_posterior}
\end{equation}

We use this framework to infer the neutral fraction at \z{7} from the sample of Lyman-break galaxies (LBGs) presented by \citet{pentericci2014}. These data are comprised of 68 LBGs with spectroscopic follow-up in legacy fields, spanning a range of UV magnitudes $-22.75 \lesssim M_\textsc{uv} \lesssim -17.8$, of which 12 had \Lya\ emission detected \citep{pentericci2014}. \citetalias{mason2018a} used the same sample for their inference, so we directly compare our new inferred neutral fraction with their result.
	
\begin{figure}
    \centering
	\includegraphics[width=\columnwidth]{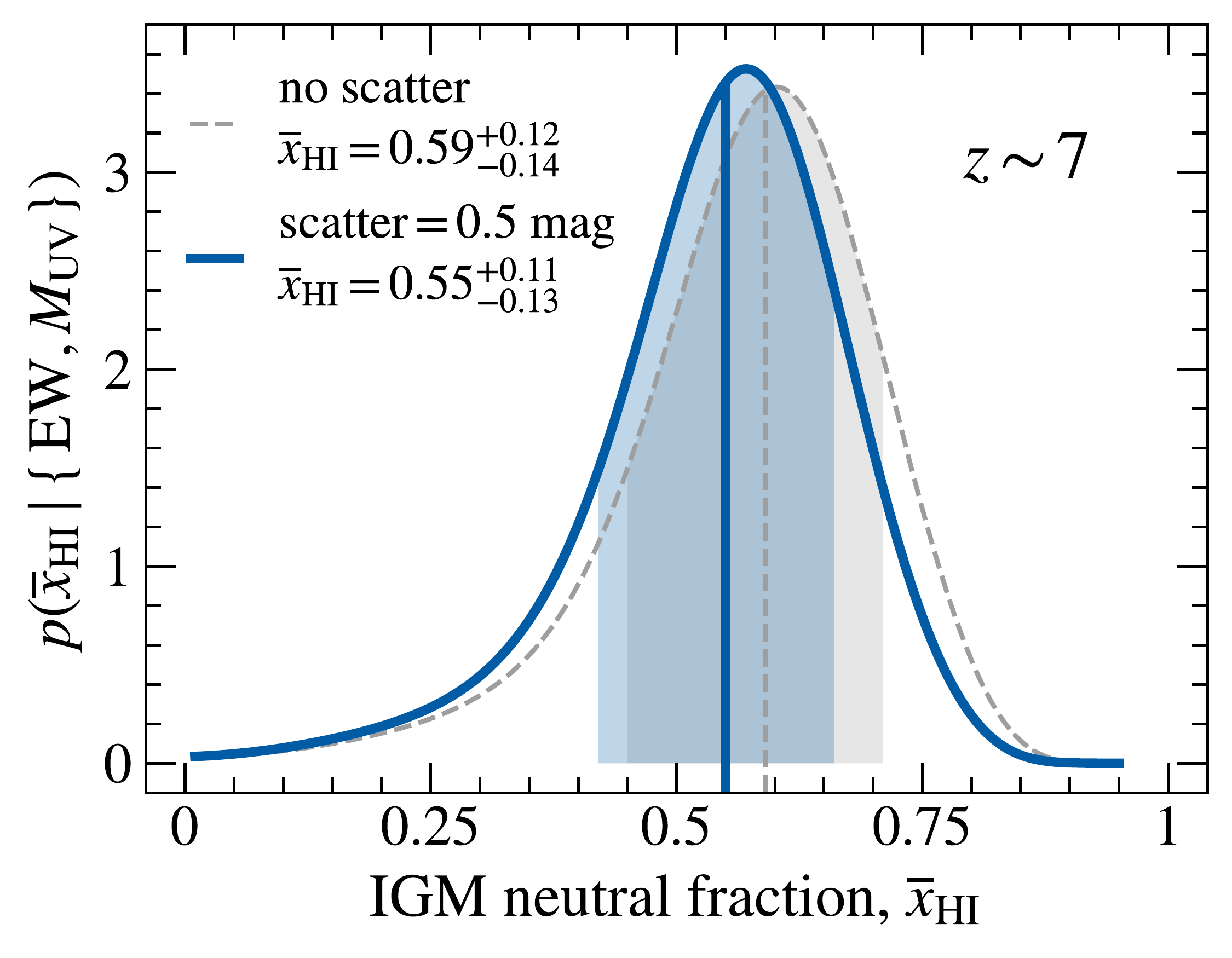}
	\caption{Posterior probability density functions for the volume-averaged fraction of neutral hydrogen in the IGM, \xHI, at \z{7} with scatter of $\sigma = 0.5$\,mag (solid blue) and without scatter (dashed grey) in the UV luminosity--halo mass relation, inferred from the \citet{pentericci2014} sample of LBGs. Vertical lines are the medians and the shaded regions are the 68 per cent credible intervals. With scatter, we infer a neutral fraction of \xHIscatter. Without scatter, the inferred neutral fraction is \xHInoscatter\ \citepalias{mason2018a}. Including scatter in the UV luminosity--halo mass relation slightly decreases the median neutral fraction inferred without scatter (i.e. we infer a more ionized IGM), but does not significantly affect the width of the posteriors.}
	\label{fig:pentericci14_xHI}
\end{figure}

In Fig.\ \ref{fig:pentericci14_xHI}, we show the posterior distributions of \xHI\ inferred from the \citet{pentericci2014} sample with and without accounting for scatter in the UV luminosity--halo mass relation. With scatter, we infer a neutral fraction of \xHIscatter\ (the errors denote the 68 per cent credible interval). Compared to the neutral fraction inferred without scatter of \xHInoscatter\ \citepalias{mason2018a}, our result is slightly lower (by $< 8$ per cent) and largely consistent within the $\sim 20 - 25$ per cent uncertainties. However, it should be noted that though the two values are consistent within the uncertainties, we have used the same data \citep{pentericci2014} and simulated dark matter haloes (see Section \ref{subsec:ISM_IGM_modelling}) to infer the neutral fraction both with and without scatter. Thus, the change in the inferred neutral fraction is due exclusively to our updated method of computing the likelihood, as described in Section \ref{subsec:CLF}.

Including scatter in the UV luminosity--halo mass relation tends to decrease \Lya\ visibility for galaxies of given UV luminosity at a fixed neutral fraction. This is especially true for UV-bright galaxies while reionization is ongoing, and the universe is neither extremely neutral nor extremely ionized; see Fig.\ \ref{fig:lya_transmission}. Thus, as scatter reduces the observed \Lya\ EW distribution, a somewhat lower neutral fraction must be invoked to explain the decline in the observed \Lya\ EWs at $z \gtrsim 6$ when scatter is included, compared to the inference without scatter.

\begin{figure}
    \centering
	\includegraphics[width=\columnwidth]{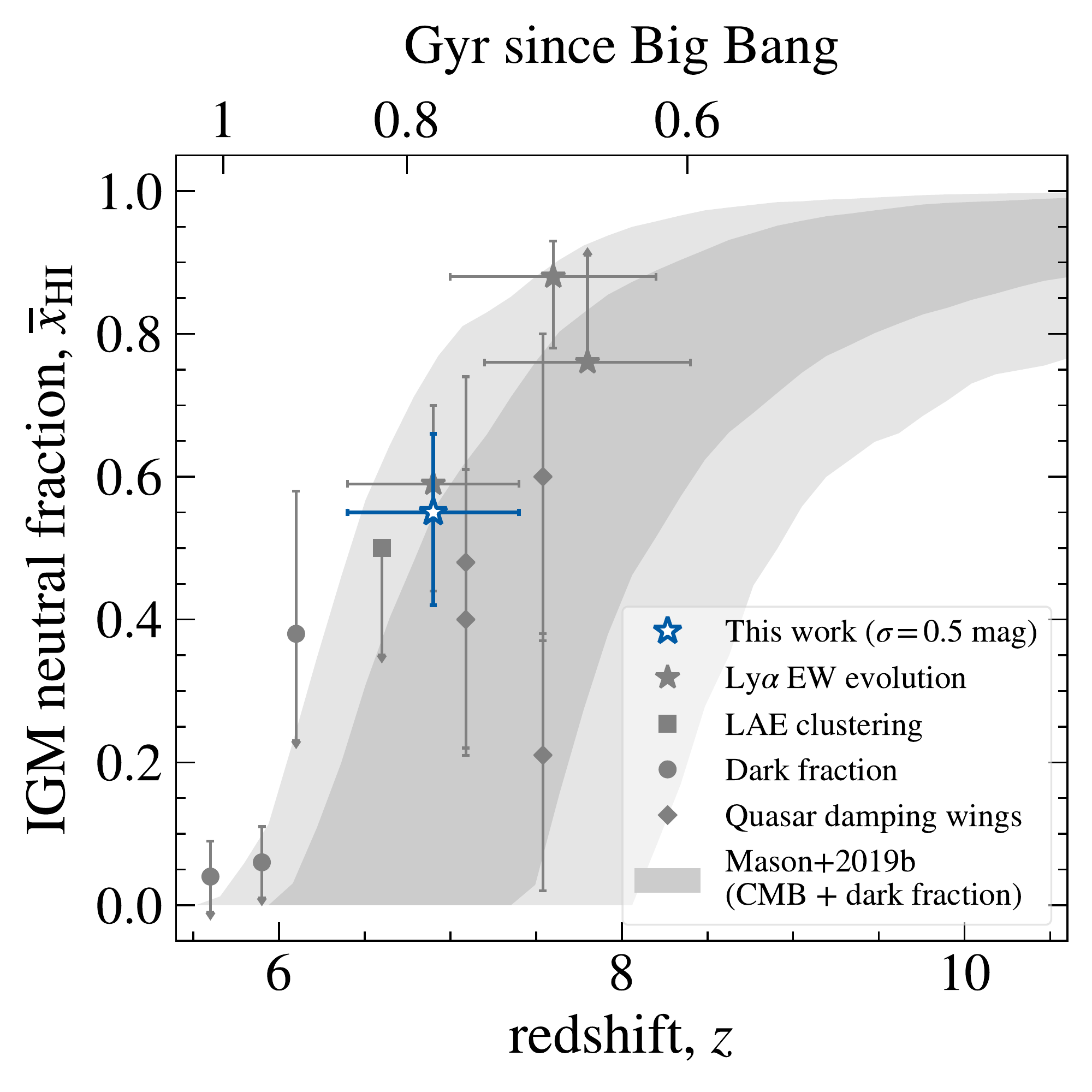}
	\caption{The redshift evolution of the volume-averaged fraction of neutral hydrogen in the IGM, \xHI. Our new measurement is shown as the blue star. We also plot measurements derived from observations of the evolving \Lya\ EW distribution without accounting for UV luminosity--halo mass relation scatter \citep[grey filled stars;][]{mason2018a, mason2019a, hoag2019}, the clustering of \Lya\ emitting galaxies \citep[square;][]{ouchi2010, sobacchi2015}, \Lya\ and Ly$\beta$ forest dark pixel fraction \citep[circles;][]{mcgreer2015}, and quasar damping wings \citep[diamonds;][]{davies2018, greig2019}. We also plot the 68 and 95 per cent credible intervals of the reionization history obtained from fitting the \citet{planck2018} CMB optical depth and dark pixel fraction constraints \citep[grey shaded regions;][]{mason2019b}.} \label{fig:EoR_history}
\end{figure}

Fig.\ \ref{fig:EoR_history} shows our result in the context of the redshift evolution of the volume-averaged fraction of neutral hydrogen in the IGM, \xHI. We show measurements derived from observations of the evolution of the \Lya\ EW distribution without scatter \citep{mason2018a, mason2019a, hoag2019}, the clustering of \Lya\ emitters \citep{ouchi2010, sobacchi2015}, \Lya\ and Ly$\beta$ forest dark pixel fraction \citep{mcgreer2015}, and quasar damping wings \citep{davies2018, greig2019}. We also show the 68 and 95 per cent credible intervals of the reionization history obtained from fitting the \citet{planck2018} CMB optical depth and dark pixel fraction constraints \citep{mason2019b}.

\subsection{Biases in the Neutral Fraction Inference} \label{subsec:biases}

Future observational campaigns with instruments such as the \textit{Wide Field Infrared Survey Telescope} may access statistical samples of \Lya\ from UV-bright LBGs at $z \gtrsim 7$ \citep{spergel2015}. Thus, we simulate luminosity-selected samples to investigate how not accounting for scatter in the UV luminosity--halo mass relation may bias the inferred neutral fraction. We simulate \Lya\ observations from galaxies, including UV luminosity--halo mass scatter, and compare the neutral fraction inferred with and without accounting for scatter.

We draw samples of galaxies from the UV luminosity function \citep{mason2015} and populate their \Lya\ EWs from the probability distribution $p(\text{EW} \mid M_\textsc{uv}, \overline{x}_{\ion{H}{i}}, \sigma)$ with scatter of $\sigma = 0.5$\,mag at a fixed neutral fraction, $\overline{x}_\ion{H}{i} = 0.49$. We also add Gaussian noise with a standard deviation of 5\,\AA\ to the \Lya\ EWs. We draw from the UV luminosity function with two limiting magnitudes (one bright, $M_\textsc{uv} \leq -21.8$, and one faint, $M_\textsc{uv} \leq -19.5$), each with two different sample sizes. We then perform the inference on the simulated samples with and without accounting for scatter in the UV luminosity--halo mass relation and compare the resulting posterior distributions (see Appendix \ref{appendix:sim_inference} for further details).

For the bright sample, without accounting for scatter, the inferred neutral fraction is biased to slightly higher values than the simulated input neutral fraction, $\overline{x}_\ion{H}{i} = 0.49$, while we recover the correct neutral fraction when scatter is accounted for. While the input neutral fraction is recovered within the uncertainty both with and without scatter for the smaller sample, the inference without scatter for the large, bright sample becomes inconsistent with the input neutral fraction as the precision on the measurement increases. See Appendix \ref{appendix:sim_inference} for further discussion.

\section{Discussion} \label{sec:discussion}

In this section, we consider caveats of our model and discuss implications of our results for future high-redshift galaxy studies. In Section~\ref{subsec:modelling_caveats}, we discuss modelling caveats. We discuss implications for the clustering of \Lya\ emitters during reionization in Section~\ref{subsec:lya_clustering}, and implications for \Lya\ in UV-bright galaxies during the EoR in Section~\ref{subsec:lya_uv_bright}. Finally, we examine the potential for constraining the scatter in the CLF in Section~\ref{subsec:constraining_scatter}.

\subsection{Modelling Caveats} \label{subsec:modelling_caveats}

In this work, we make several significant assumptions. First, following \citetalias{mason2018a}, we assume that the intrinsic emitted (i.e. after processing the ISM) \Lya\ EW distribution at \z{7} is the same as the observed \Lya\ EW distribution at \z{6} (see Section~\ref{subsec:ISM_IGM_modelling}). This is likely a simplification, as trends at lower redshifts show an evolution towards higher \Lya\ EWs with redshift as dust fraction decreases \citep{hayes2011}. However, this assumption also likely results in a lower inferred neutral fraction than if the \Lya\ EW distribution at \z{7} is not the same as the observed \Lya\ EW distribution at \z{6}. If the emitted \Lya\ EW distribution evolves significantly in the short time between \z{7} and \z{6} ($\lesssim 200$\,Myr), it will probably continue to follow the lower-redshift trend and have higher \Lya\ EWs at \z{7}. This would require a larger neutral fraction than we infer to suppress our modelled \Lya\ EW distribution to the observed \Lya\ EW distribution. However, we note that the evolution of the intrinsic \Lya\ EW distribution is not fully understood: \citet{debarros2017} found tentative evidence for decreasing \Lya\ EWs between $z \sim 5 - 6$ which could be due to intrinsic evolution, or an indication there is still neutral gas in some IGM sightlines at $z \sim 5-6$ \citep[as suggested by an extended Gunn-Peterson trough at $z \sim 5.5$;][]{becker2015,keating2019}.

We also consider only the distribution of galaxies within the IGM and do not model any change in the IGM topology when scatter in the UV luminosity--halo mass relation is included. This is reasonable if low-mass, faint galaxies dominate reionization, as including scatter has a smaller impact on these galaxies. However, if brighter galaxies do contribute significantly to reionization, scatter could change the IGM topology. For example, UV-bright galaxies in low-mass haloes may be bright due to a burst of star formation, and ionizing photons produced in that burst could more easily escape from a low-mass halo than from a high-mass halo due to the smaller potential well \citep[e.g.][]{wise2014, pardekooper2015}. Hence, UV-bright galaxies in low-mass haloes could reside in larger ionized regions than we consider, particularly in the earliest stages of reionization; in later stages, the EoR topology depends primarily on the cumulative ionizing contribution from all sources. Thus, here we model the maximum expected difference in \Lya\ visibility from the UV luminosity--halo mass relation without scatter. Accounting for the changing IGM topology would likely increase \Lya\ visibility, and therefore the IGM neutral fraction, compared to our results in this work.

Future improvements to the model to account for these assumptions would, on average, most likely result in higher observed \Lya\ EWs at fixed neutral fraction (particularly for UV-bright galaxies), and therefore we would infer a higher neutral fraction than we find in this work.

Finally, this work assumes that the evolution of \Lya\ visibility between \z{7} and \z{6} is due to the evolution of the \Lya\ damping wing optical depth of the diffuse neutral component of the IGM. We do not model redshift evolution of \Lya\ transmission in the ionized component of the IGM and CGM at fixed halo mass, which can depend strongly on the velocity offset from the systemic redshift of the source \citep[e.g.][]{dijkstra2011, laursen2011, choudhury2015}. This assumption is reasonable if the \Lya\ line profiles emerging from LBGs are dominated by scattering off outflows \citep[e.g.,][]{verhamme2008}, which produces high $\Delta v$ \citep[as at lower redshifts, e.g.][]{steidel2010, hashimoto2013, shibuya2014}. However, if \Lya\ velocity offsets decrease significantly at fixed halo mass at $z>6$, decrease in the transmission of \Lya\ through the ionized IGM and CGM may be significant \citep{dijkstra2011,choudhury2015}, and lessen the requirement for a rapid evolution in \xHI. We will explore the impact on \xHI\ inferences of redshift-dependent velocity offsets and the evolution of resonant absorption in the local ionized region around the source, $\tau_{\ion{H}{ii}}$, in future work.

\subsection{Clustering of \texorpdfstring{L\MakeLowercase{y}$\alpha$}{Lya} Emitters} \label{subsec:lya_clustering}

In this work, we examine the evolution of the observed \Lya\ EW distribution between \z{6} and \z{7} due to absorption by neutral hydrogen in the IGM during the EoR. However, the clustering signal of \Lya\ emitters (LAEs, selected on their \Lya\ flux rather than UV continuum) is also sensitive to attenuation by neutral hydrogen; \Lya\ emission is preferentially transmitted through the large ionized bubbles surrounding clusters of LAEs, therefore increasing the clustering signal \citep[e.g.][]{sobacchi2015, weinberger2019}.

Recent measurements of the LAE angular correlation function by \citet{ouchi2018} show no significant evolution from \z{5.7} to \z{6.6}, a somewhat unexpected result given the expected amplification of the clustering signal during reionization. Now, with scatter in the UV luminosity--halo mass relation, galaxies are hosted by lower mass halos on average, thus the probability of a low-mass halo hosting an LAE increases, implying some weakening of the LAE clustering signal. However, this depends on the relationship between \Muv\ and \Lya\ flux: if LAEs are biased towards UV-luminous galaxies, the decrease in clustering due to UV luminosity--halo mass scatter may be negligible. We leave quantitative study of the effects of scatter in the UV luminosity--halo mass relation on LAE clustering for future work.

\subsection{\texorpdfstring{L\MakeLowercase{y}$\alpha$}{Lya} in UV-bright Galaxies During the EoR} \label{subsec:lya_uv_bright}

Recent detections of \Lya\ at $z \gtrsim 7.5$ from UV-bright galaxies \citep[e.g.][]{oesch2015, zitrin2015, roberts-borsani2016, stark2017} are somewhat surprising, since at lower redshifts, UV-bright galaxies are unlikely to have strong \Lya\ emission \citep[e.g.][]{stark2010}. The inhomogeneous nature of reionization likely accounts for some of this apparent increase in \Lya\ from UV-bright galaxies, due to higher \Lya\ transmission through the reionizing IGM from bright galaxies residing in large ionized bubbles (Fig.\ \ref{fig:lya_transmission}). However, \citet{mason2018b} found that merely increased \Lya\ transmission through the IGM for extremely UV-bright galaxies ($M_\textsc{uv} \lesssim -22$) is insufficient to explain the evolution of the \Lya\ fraction for these galaxies between $6 \lesssim z \lesssim 8$ \citep{stark2017}. This suggests that the emitted \Lya\ EWs from these bright galaxies are higher than expected. In this work, we find a decrease in \Lya\ transmission for UV-bright galaxies from the models of \citetalias{mason2018a} (Fig.\ \ref{fig:lya_transmission}), implying that the emitted \Lya\ EWs of the \citet{roberts-borsani2016, stark2017} sample are even higher than previously required by \cite{mason2018b} to explain their observed emission. The implied high intrinsic EW of this sample, relative to other galaxies of a similar UV luminosity, is likely due to atypical intrinsic galactic properties, selection effects, and/or that these sources have hard ionizing spectra which may have highly ionized their surroundings, reducing the optical depth to \Lya\ in the ionized IGM \citep[see][for further discussion]{stark2017, mason2018b}.

Our analysis in Section \ref{subsec:biases} and Appendix \ref{appendix:sim_inference} indicates that using UV-bright galaxies ($M_\textsc{uv} \lesssim -22$) to constrain \xHI\ \textit{without} accounting for UV luminosity--halo mass scatter can lead to biased measurements of \xHI. As the magnitude of the scatter itself is uncertain (see below), extremely UV-bright galaxies should only be used as probes of the IGM if the impact of UV luminosity--halo mass scatter is marginalised over. However, it should be noted that the scatter is only a significant systematic uncertainty for large samples ($N \gtrsim 1000$) of UV-bright galaxies. For samples that include fainter galaxies, the neutral fraction is recovered well with or without scatter, even for samples as large as $N = 1000$.

\subsection{Constraining UV Luminosity--Halo Mass Scatter} \label{subsec:constraining_scatter}

The degree of scatter in the CLF is a free parameter, and we chose $\sigma = 0.5$\,mag as our fiducial value, since \citetalias{ren2019} found that this scatter is consistent with observations at $z \sim 6 - 10$. 

To maintain consistency with observations, \citetalias{ren2019} also applied a critical flattening threshold to the median UV magnitude--halo mass relation based on either mass or luminosity, which can be related to feedback processes by active galactic nuclei. \citetalias{ren2019} explored alternative values of $\sigma$ with both mass and luminosity flattening thresholds, and found that for high-scatter cases ($\sigma > 0.5$\,mag), scatter and flattening criteria were degenerate when using only the UV luminosity function to discriminate between cases. This degeneracy arises because the CLF does not fully recover the brightest end of the observed luminosity function\footnote{A smaller dispersion with a critical mass flattening threshold is difficult to distinguish from a larger dispersion combined with a critical luminosity flattening threshold. A critical mass threshold tends to flatten the median UV luminosity--halo mass relation at brighter luminosities than a critical luminosity threshold does \citepalias{ren2019}, so a critical mass threshold with smaller dispersion predicts approximately the same abundance of bright objects as a critical luminosity threshold with larger dispersion.} (see \citetalias{ren2019} for further discussion).

\citetalias{ren2019} suggest that an independent constraint on the scatter, such as measurements of the local clustering strength around UV-bright galaxies, would break this degeneracy. Alternatively, a precise measurement of the EoR timeline could constrain the degree of the scatter. The transmission of \Lya\ through the IGM is sensitive to variations in the column density of neutral hydrogen along the line of sight, which in turn depends on the distribution of the ionizing sources (i.e. galaxies). Thus, our model discriminates between larger and smaller scatter, though it is largely insensitive to the flattening threshold at \z{7}. However, determining the correct scatter would require a precision measurement of the EoR timeline by another method (for example, via measurements of the redshifted 21\,cm line from neutral hydrogen).

\section{Summary and Conclusions} \label{sec:conclusion}

We have updated a model of galaxy properties during the EoR to include physically-motivated scatter in the relation between galaxy UV luminosity and dark matter halo mass. We have modelled \Lya\ visibility during the EoR, inferred the fraction of neutral hydrogen in the IGM at \z{7} with the \citet{pentericci2014} sample of LBGs, and compared with results from models without scatter in the UV luminosity--halo mass relation \citepalias{mason2018a}. Our primary conclusions are as follows:

\begin{enumerate}
	\item Scatter in the galaxy UV luminosity to halo mass relation alters the distribution of galaxies within the IGM, and thus impacts \Lya\ visibility from galaxies.
	\item Scatter reduces \Lya\ visibility for very UV-luminous objects. This is due to the increased probability of a UV-bright galaxy residing in a lower-mass halo surrounded by a smaller ionized region, thus increasing the probability of absorption by neutral hydrogen along the line of sight. This effect is less significant for fainter galaxies, as they are much more numerous (so introducing outliers has a smaller effect) and already reside in low-mass haloes surrounded by small ionized regions.
	\item Scatter slightly decreases the median neutral fraction inferred at \z{7} from the \citet{pentericci2014} sample of 68 LBGs compared to the inference without accounting for scatter. With scatter, we infer a neutral fraction of \xHIscatter, compared to the neutral fraction inferred without scatter of \xHInoscatter, \citepalias{mason2018a}. However, scatter does not impact the width of the probability distribution for the neutral fraction.
	\item Using samples of UV-bright galaxies to constrain \xHI\ without accounting for scatter in the UV luminosity--halo mass relation leads to overestimates of \xHI. The discrepancy becomes more significant for larger sample sizes.
\end{enumerate}

More refinements to our model, such as including the evolution of the intrinsic \Lya\ EW distribution, accounting for the changing IGM during the EoR, modelling the evolution of $\Delta v$ with redshift, and fully constraining the scatter in the galaxy UV luminosity--halo mass relation, will enable us to further understand the nature of scatter and its impact on \Lya\ visibility during reionization. Along with observational campaigns with, e.g. the \textit{James Webb Space Telescope}, we will be able to place precision constraints on the timeline and topology of the EoR and understand the properties of the sources that drive reionization.

\section*{Acknowledgements}

The authors thank Charlie Conroy for useful discussions, and Matt Ashby and Jonathan McDowell for providing comments on an early draft. We thank Stephane de Barros for providing the observed \z{6} \Lya\ EW sample from \citet{debarros2017}.

This work is based on data obtained from ESO programs 190.A-0685 and 088.A-1013. L.R.W. acknowledges the Smithsonian Astrophysical Observatory REU program, which is funded in part by the NSF REU and Department of Defense ASSURE programs under NSF Grant no.\ AST-1852268, and by the Smithsonian Institution. C.A.M. acknowledges support by NASA Headquarters through the NASA Hubble Fellowship grant HST-HF2-51413.001-A awarded by the Space Telescope Science Institute, which is operated by the Association of Universities for Research in Astronomy, Inc., for NASA, under contract NAS5-26555. K.R. acknowledges support through the Research Training Program Scholarship from the Australian Government. This work was supported in part by the NSF grant: COLLABORATIVE RESEARCH: The Final Frontier: Spectroscopic Probes of Galaxies at the Epoch of Reionization (AST-1815458, AST-1810822).

This work made use of the following software packages: NumPy \citep{oliphant2006, vanderWalt2011}, SciPy \citep{jones2001}, IPython \citep{perez2007}, Matplotlib \citep{hunter2007}, Astropy \citep{astropy2013, astropy2018}, and \textsc{hmf} \citep{murray2013}.



\bibliographystyle{mnras}
\bibliography{refs}

\begin{thebibliography}{}
\makeatletter
\relax
\def\mn@urlcharsother{\let\do\@makeother \do\$\do\&\do\#\do\^\do\_\do\%\do\~}
\def\mn@doi{\begingroup\mn@urlcharsother \@ifnextchar [ {\mn@doi@}
  {\mn@doi@[]}}
\def\mn@doi@[#1]#2{\def\@tempa{#1}\ifx\@tempa\@empty \href
  {http://dx.doi.org/#2} {doi:#2}\else \href {http://dx.doi.org/#2} {#1}\fi
  \endgroup}
\def\mn@eprint#1#2{\mn@eprint@#1:#2::\@nil}
\def\mn@eprint@arXiv#1{\href {http://arxiv.org/abs/#1} {{\tt arXiv:#1}}}
\def\mn@eprint@dblp#1{\href {http://dblp.uni-trier.de/rec/bibtex/#1.xml}
  {dblp:#1}}
\def\mn@eprint@#1:#2:#3:#4\@nil{\def\@tempa {#1}\def\@tempb {#2}\def\@tempc
  {#3}\ifx \@tempc \@empty \let \@tempc \@tempb \let \@tempb \@tempa \fi \ifx
  \@tempb \@empty \def\@tempb {arXiv}\fi \@ifundefined
  {mn@eprint@\@tempb}{\@tempb:\@tempc}{\expandafter \expandafter \csname
  mn@eprint@\@tempb\endcsname \expandafter{\@tempc}}}

\bibitem[\protect\citeauthoryear{{Astropy Collaboration} et~al.,}{{Astropy
  Collaboration} et~al.}{2013}]{astropy2013}
{Astropy Collaboration} et~al., 2013, \mn@doi [\aap]
  {10.1051/0004-6361/201322068}, \href
  {https://ui.adsabs.harvard.edu/abs/2013A&A...558A..33A} {558, A33}

\bibitem[\protect\citeauthoryear{{Astropy Collaboration} et~al.,}{{Astropy
  Collaboration} et~al.}{2018}]{astropy2018}
{Astropy Collaboration} et~al., 2018, \mn@doi [\aj] {10.3847/1538-3881/aabc4f},
  \href {https://ui.adsabs.harvard.edu/abs/2018AJ....156..123A} {156, 123}

\bibitem[\protect\citeauthoryear{{Ba{\~n}ados} et~al.,}{{Ba{\~n}ados}
  et~al.}{2018}]{banados2018}
{Ba{\~n}ados} E.,  et~al., 2018, \mn@doi [\nat] {10.1038/nature25180}, \href
  {https://ui.adsabs.harvard.edu/abs/2018Natur.553..473B} {553, 473}

\bibitem[\protect\citeauthoryear{{Barkana} \& {Loeb}}{{Barkana} \&
  {Loeb}}{2001}]{barkana2001}
{Barkana} R.,  {Loeb} A.,  2001, \mn@doi [\physrep]
  {10.1016/S0370-1573(01)00019-9}, \href
  {https://ui.adsabs.harvard.edu/abs/2001PhR...349..125B} {349, 125}

\bibitem[\protect\citeauthoryear{{Becker}, {Bolton}, {Madau}, {Pettini},
  {Ryan-Weber}  \& {Venemans}}{{Becker} et~al.}{2015}]{becker2015}
{Becker} G.~D.,  {Bolton} J.~S.,  {Madau} P.,  {Pettini} M.,  {Ryan-Weber}
  E.~V.,   {Venemans} B.~P.,  2015, \mn@doi [\mnras] {10.1093/mnras/stu2646},
  \href {https://ui.adsabs.harvard.edu/abs/2015MNRAS.447.3402B} {447, 3402}

\bibitem[\protect\citeauthoryear{{Behroozi}, {Conroy}  \&
  {Wechsler}}{{Behroozi} et~al.}{2010}]{behroozi2010}
{Behroozi} P.~S.,  {Conroy} C.,   {Wechsler} R.~H.,  2010, \mn@doi [\apj]
  {10.1088/0004-637X/717/1/379}, \href
  {https://ui.adsabs.harvard.edu/abs/2010ApJ...717..379B} {717, 379}

\bibitem[\protect\citeauthoryear{{Behroozi}, {Wechsler}  \&
  {Conroy}}{{Behroozi} et~al.}{2013}]{behroozi2013}
{Behroozi} P.~S.,  {Wechsler} R.~H.,   {Conroy} C.,  2013, \mn@doi [\apj]
  {10.1088/0004-637X/770/1/57}, \href
  {https://ui.adsabs.harvard.edu/abs/2013ApJ...770...57B} {770, 57}

\bibitem[\protect\citeauthoryear{{Bouwens} et~al.,}{{Bouwens}
  et~al.}{2015}]{bouwens2015}
{Bouwens} R.~J.,  et~al., 2015, \mn@doi [\apj] {10.1088/0004-637X/803/1/34},
  \href {https://ui.adsabs.harvard.edu/abs/2015ApJ...803...34B} {803, 34}

\bibitem[\protect\citeauthoryear{{Cassata} et~al.,}{{Cassata}
  et~al.}{2015}]{cassata2015}
{Cassata} P.,  et~al., 2015, \mn@doi [\aap] {10.1051/0004-6361/201423824},
  \href {https://ui.adsabs.harvard.edu/abs/2015A&A...573A..24C} {573, A24}

\bibitem[\protect\citeauthoryear{{Choudhury}, {Puchwein}, {Haehnelt}  \&
  {Bolton}}{{Choudhury} et~al.}{2015}]{choudhury2015}
{Choudhury} T.~R.,  {Puchwein} E.,  {Haehnelt} M.~G.,   {Bolton} J.~S.,  2015,
  \mn@doi [\mnras] {10.1093/mnras/stv1250}, \href
  {https://ui.adsabs.harvard.edu/abs/2015MNRAS.452..261C} {452, 261}

\bibitem[\protect\citeauthoryear{{Curtis-Lake} et~al.,}{{Curtis-Lake}
  et~al.}{2012}]{curtis-lake2012}
{Curtis-Lake} E.,  et~al., 2012, \mn@doi [\mnras]
  {10.1111/j.1365-2966.2012.20720.x}, \href
  {https://ui.adsabs.harvard.edu/abs/2012MNRAS.422.1425C} {422, 1425}

\bibitem[\protect\citeauthoryear{{Davies} et~al.,}{{Davies}
  et~al.}{2018}]{davies2018}
{Davies} F.~B.,  et~al., 2018, \mn@doi [\apj] {10.3847/1538-4357/aad6dc}, \href
  {https://ui.adsabs.harvard.edu/abs/2018ApJ...864..142D} {864, 142}

\bibitem[\protect\citeauthoryear{{Dayal} \& {Ferrara}}{{Dayal} \&
  {Ferrara}}{2018}]{dayal2018}
{Dayal} P.,  {Ferrara} A.,  2018, \mn@doi [\physrep]
  {10.1016/j.physrep.2018.10.002}, \href
  {https://ui.adsabs.harvard.edu/abs/2018PhR...780....1D} {780, 1}

\bibitem[\protect\citeauthoryear{{De Barros} et~al.,}{{De Barros}
  et~al.}{2017}]{debarros2017}
{De Barros} S.,  et~al., 2017, \mn@doi [\aap] {10.1051/0004-6361/201731476},
  \href {https://ui.adsabs.harvard.edu/abs/2017A&A...608A.123D} {608, A123}

\bibitem[\protect\citeauthoryear{{Dijkstra}}{{Dijkstra}}{2014}]{dijkstra2014}
{Dijkstra} M.,  2014, \mn@doi [\pasa] {10.1017/pasa.2014.33}, \href
  {https://ui.adsabs.harvard.edu/abs/2014PASA...31...40D} {31, e040}

\bibitem[\protect\citeauthoryear{{Dijkstra}, {Lidz}  \& {Wyithe}}{{Dijkstra}
  et~al.}{2007}]{dijkstra2007}
{Dijkstra} M.,  {Lidz} A.,   {Wyithe} J. S.~B.,  2007, \mn@doi [\mnras]
  {10.1111/j.1365-2966.2007.11666.x}, \href
  {https://ui.adsabs.harvard.edu/abs/2007MNRAS.377.1175D} {377, 1175}

\bibitem[\protect\citeauthoryear{{Dijkstra}, {Mesinger}  \&
  {Wyithe}}{{Dijkstra} et~al.}{2011}]{dijkstra2011}
{Dijkstra} M.,  {Mesinger} A.,   {Wyithe} J. S.~B.,  2011, \mn@doi [\mnras]
  {10.1111/j.1365-2966.2011.18530.x}, \href
  {https://ui.adsabs.harvard.edu/abs/2011MNRAS.414.2139D} {414, 2139}

\bibitem[\protect\citeauthoryear{{Fan}, {Carilli}  \& {Keating}}{{Fan}
  et~al.}{2006}]{fan2006}
{Fan} X.,  {Carilli} C.~L.,   {Keating} B.,  2006, \mn@doi [\araa]
  {10.1146/annurev.astro.44.051905.092514}, \href
  {https://ui.adsabs.harvard.edu/abs/2006ARA&A..44..415F} {44, 415}

\bibitem[\protect\citeauthoryear{{Finkelstein} et~al.,}{{Finkelstein}
  et~al.}{2015}]{finkelstein2015}
{Finkelstein} S.~L.,  et~al., 2015, \mn@doi [\apj]
  {10.1088/0004-637X/814/2/95}, \href
  {https://ui.adsabs.harvard.edu/abs/2015ApJ...814...95F} {814, 95}

\bibitem[\protect\citeauthoryear{{Furlanetto}, {Zaldarriaga}  \&
  {Hernquist}}{{Furlanetto} et~al.}{2004}]{furlanetto2004}
{Furlanetto} S.~R.,  {Zaldarriaga} M.,   {Hernquist} L.,  2004, \mn@doi [\apj]
  {10.1086/423025}, \href
  {https://ui.adsabs.harvard.edu/abs/2004ApJ...613....1F} {613, 1}

\bibitem[\protect\citeauthoryear{{Furlanetto}, {Zaldarriaga}  \&
  {Hernquist}}{{Furlanetto} et~al.}{2006}]{furlanetto2006}
{Furlanetto} S.~R.,  {Zaldarriaga} M.,   {Hernquist} L.,  2006, \mn@doi
  [\mnras] {10.1111/j.1365-2966.2005.09785.x}, \href
  {https://ui.adsabs.harvard.edu/abs/2006MNRAS.365.1012F} {365, 1012}

\bibitem[\protect\citeauthoryear{{Greig}, {Mesinger}, {Haiman}  \&
  {Simcoe}}{{Greig} et~al.}{2017}]{greig2017}
{Greig} B.,  {Mesinger} A.,  {Haiman} Z.,   {Simcoe} R.~A.,  2017, \mn@doi
  [\mnras] {10.1093/mnras/stw3351}, \href
  {https://ui.adsabs.harvard.edu/abs/2017MNRAS.466.4239G} {466, 4239}

\bibitem[\protect\citeauthoryear{{Greig}, {Mesinger}  \& {Ba{\~n}ados}}{{Greig}
  et~al.}{2019}]{greig2019}
{Greig} B.,  {Mesinger} A.,   {Ba{\~n}ados} E.,  2019, \mn@doi [\mnras]
  {10.1093/mnras/stz230}, \href
  {https://ui.adsabs.harvard.edu/abs/2019MNRAS.484.5094G} {484, 5094}

\bibitem[\protect\citeauthoryear{{Gunn} \& {Peterson}}{{Gunn} \&
  {Peterson}}{1965}]{gunn1965}
{Gunn} J.~E.,  {Peterson} B.~A.,  1965, \mn@doi [\apj] {10.1086/148444}, \href
  {https://ui.adsabs.harvard.edu/abs/1965ApJ...142.1633G} {142, 1633}

\bibitem[\protect\citeauthoryear{{Haiman} \& {Spaans}}{{Haiman} \&
  {Spaans}}{1999}]{haiman1999}
{Haiman} Z.,  {Spaans} M.,  1999, \mn@doi [\apj] {10.1086/307276}, \href
  {https://ui.adsabs.harvard.edu/abs/1999ApJ...518..138H} {518, 138}

\bibitem[\protect\citeauthoryear{{Hashimoto}, {Ouchi}, {Shimasaku}, {Ono},
  {Nakajima}, {Rauch}, {Lee}  \& {Okamura}}{{Hashimoto}
  et~al.}{2013}]{hashimoto2013}
{Hashimoto} T.,  {Ouchi} M.,  {Shimasaku} K.,  {Ono} Y.,  {Nakajima} K.,
  {Rauch} M.,  {Lee} J.,   {Okamura} S.,  2013, \mn@doi [\apj]
  {10.1088/0004-637X/765/1/70}, \href
  {https://ui.adsabs.harvard.edu/abs/2013ApJ...765...70H} {765, 70}

\bibitem[\protect\citeauthoryear{{Hayes}, {Schaerer}, {{\"O}stlin},
  {Mas-Hesse}, {Atek}  \& {Kunth}}{{Hayes} et~al.}{2011}]{hayes2011}
{Hayes} M.,  {Schaerer} D.,  {{\"O}stlin} G.,  {Mas-Hesse} J.~M.,  {Atek} H.,
  {Kunth} D.,  2011, \mn@doi [\apj] {10.1088/0004-637X/730/1/8}, \href
  {https://ui.adsabs.harvard.edu/abs/2011ApJ...730....8H} {730, 8}

\bibitem[\protect\citeauthoryear{{Hoag} et~al.,}{{Hoag}
  et~al.}{2019}]{hoag2019}
{Hoag} A.,  et~al., 2019, \mn@doi [\apj] {10.3847/1538-4357/ab1de7}, \href
  {https://ui.adsabs.harvard.edu/abs/2019ApJ...878...12H} {878, 12}

\bibitem[\protect\citeauthoryear{{Hunter}}{{Hunter}}{2007}]{hunter2007}
{Hunter} J.~D.,  2007, \mn@doi [Computing in Science and Engineering]
  {10.1109/MCSE.2007.55}, \href
  {https://ui.adsabs.harvard.edu/abs/2007CSE.....9...90H} {9, 90}

\bibitem[\protect\citeauthoryear{{Jensen}, {Hayes}, {Iliev}, {Laursen},
  {Mellema}  \& {Zackrisson}}{{Jensen} et~al.}{2014}]{jensen2014}
{Jensen} H.,  {Hayes} M.,  {Iliev} I.~T.,  {Laursen} P.,  {Mellema} G.,
  {Zackrisson} E.,  2014, \mn@doi [\mnras] {10.1093/mnras/stu1600}, \href
  {https://ui.adsabs.harvard.edu/abs/2014MNRAS.444.2114J} {444, 2114}

\bibitem[\protect\citeauthoryear{{Jones}, {Oliphant}, {Peterson}
  et~al.}{{Jones} et~al.}{2001}]{jones2001}
{Jones} E.,  {Oliphant} T.,  {Peterson} P.,   et~al., 2001, {SciPy: Open source
  scientific tools for Python}, \url {http://www.scipy.org/}

\bibitem[\protect\citeauthoryear{{Keating}, {Weinberger}, {Kulkarni},
  {Haehnelt}, {Chardin}  \& {Aubert}}{{Keating} et~al.}{2019}]{keating2019}
{Keating} L.~C.,  {Weinberger} L.~H.,  {Kulkarni} G.,  {Haehnelt} M.~G.,
  {Chardin} J.,   {Aubert} D.,  2019, arXiv e-prints, \href
  {https://ui.adsabs.harvard.edu/abs/2019arXiv190512640K} {p. arXiv:1905.12640}

\bibitem[\protect\citeauthoryear{{Komatsu} et~al.,}{{Komatsu}
  et~al.}{2011}]{komatsu2011}
{Komatsu} E.,  et~al., 2011, \mn@doi [\apjs] {10.1088/0067-0049/192/2/18},
  \href {https://ui.adsabs.harvard.edu/abs/2011ApJS..192...18K} {192, 18}

\bibitem[\protect\citeauthoryear{{Kulkarni}, {Keating}, {Haehnelt}, {Bosman},
  {Puchwein}, {Chardin}  \& {Aubert}}{{Kulkarni} et~al.}{2019a}]{kulkarni2019a}
{Kulkarni} G.,  {Keating} L.~C.,  {Haehnelt} M.~G.,  {Bosman} S. E.~I.,
  {Puchwein} E.,  {Chardin} J.,   {Aubert} D.,  2019a, \mn@doi [\mnras]
  {10.1093/mnrasl/slz025}, \href
  {https://ui.adsabs.harvard.edu/abs/2019MNRAS.485L..24K} {485, L24}

\bibitem[\protect\citeauthoryear{{Kulkarni}, {Worseck}  \&
  {Hennawi}}{{Kulkarni} et~al.}{2019b}]{kulkarni2019b}
{Kulkarni} G.,  {Worseck} G.,   {Hennawi} J.~F.,  2019b, \mn@doi [\mnras]
  {10.1093/mnras/stz1493}, \href
  {https://ui.adsabs.harvard.edu/abs/2019MNRAS.488.1035K} {488, 1035}

\bibitem[\protect\citeauthoryear{{Kullback}}{{Kullback}}{1968}]{kullback1968}
{Kullback} S.,  1968, {Information theory and statistics}.
Dover Publications

\bibitem[\protect\citeauthoryear{{Laursen}, {Sommer-Larsen}  \&
  {Razoumov}}{{Laursen} et~al.}{2011}]{laursen2011}
{Laursen} P.,  {Sommer-Larsen} J.,   {Razoumov} A.~O.,  2011, \mn@doi [\apj]
  {10.1088/0004-637X/728/1/52}, \href
  {https://ui.adsabs.harvard.edu/abs/2011ApJ...728...52L} {728, 52}

\bibitem[\protect\citeauthoryear{{Lidz}, {McQuinn}, {Zaldarriaga}, {Hernquist}
  \& {Dutta}}{{Lidz} et~al.}{2007}]{lidz2007}
{Lidz} A.,  {McQuinn} M.,  {Zaldarriaga} M.,  {Hernquist} L.,   {Dutta} S.,
  2007, \mn@doi [\apj] {10.1086/521974}, \href
  {https://ui.adsabs.harvard.edu/abs/2007ApJ...670...39L} {670, 39}

\bibitem[\protect\citeauthoryear{{Malhotra} \& {Rhoads}}{{Malhotra} \&
  {Rhoads}}{2004}]{malhotra2004}
{Malhotra} S.,  {Rhoads} J.~E.,  2004, \mn@doi [\apj] {10.1086/427182}, \href
  {https://ui.adsabs.harvard.edu/abs/2004ApJ...617L...5M} {617, L5}

\bibitem[\protect\citeauthoryear{{Mason}, {Trenti}  \& {Treu}}{{Mason}
  et~al.}{2015}]{mason2015}
{Mason} C.~A.,  {Trenti} M.,   {Treu} T.,  2015, \mn@doi [\apj]
  {10.1088/0004-637X/813/1/21}, \href
  {https://ui.adsabs.harvard.edu/abs/2015ApJ...813...21M} {813, 21}

\bibitem[\protect\citeauthoryear{{Mason}, {Treu}, {Dijkstra}, {Mesinger},
  {Trenti}, {Pentericci}, {de Barros}  \& {Vanzella}}{{Mason}
  et~al.}{2018a}]{mason2018a}
{Mason} C.~A.,  {Treu} T.,  {Dijkstra} M.,  {Mesinger} A.,  {Trenti} M.,
  {Pentericci} L.,  {de Barros} S.,   {Vanzella} E.,  2018a, \mn@doi [\apj]
  {10.3847/1538-4357/aab0a7}, \href
  {https://ui.adsabs.harvard.edu/abs/2018ApJ...856....2M} {856, 2}

\bibitem[\protect\citeauthoryear{{Mason} et~al.,}{{Mason}
  et~al.}{2018b}]{mason2018b}
{Mason} C.~A.,  et~al., 2018b, \mn@doi [\apj] {10.3847/2041-8213/aabbab}, \href
  {https://ui.adsabs.harvard.edu/abs/2018ApJ...857L..11M} {857, L11}

\bibitem[\protect\citeauthoryear{{Mason} et~al.,}{{Mason}
  et~al.}{2019a}]{mason2019a}
{Mason} C.~A.,  et~al., 2019a, \mn@doi [\mnras] {10.1093/mnras/stz632}, \href
  {https://ui.adsabs.harvard.edu/abs/2019MNRAS.485.3947M} {485, 3947}

\bibitem[\protect\citeauthoryear{{Mason}, {Naidu}, {Tacchella}  \&
  {Leja}}{{Mason} et~al.}{2019b}]{mason2019b}
{Mason} C.~A.,  {Naidu} R.~P.,  {Tacchella} S.,   {Leja} J.,  2019b, \mn@doi
  [\mnras] {10.1093/mnras/stz2291}, \href
  {https://ui.adsabs.harvard.edu/abs/2019MNRAS.489.2669M} {489, 2669}

\bibitem[\protect\citeauthoryear{{McGreer}, {Mesinger}  \&
  {D'Odorico}}{{McGreer} et~al.}{2015}]{mcgreer2015}
{McGreer} I.~D.,  {Mesinger} A.,   {D'Odorico} V.,  2015, \mn@doi [\mnras]
  {10.1093/mnras/stu2449}, \href
  {https://ui.adsabs.harvard.edu/abs/2015MNRAS.447..499M} {447, 499}

\bibitem[\protect\citeauthoryear{{McQuinn}, {Lidz}, {Zahn}, {Dutta},
  {Hernquist}  \& {Zaldarriaga}}{{McQuinn} et~al.}{2007}]{mcquinn2007}
{McQuinn} M.,  {Lidz} A.,  {Zahn} O.,  {Dutta} S.,  {Hernquist} L.,
  {Zaldarriaga} M.,  2007, \mn@doi [\mnras] {10.1111/j.1365-2966.2007.11489.x},
  \href {https://ui.adsabs.harvard.edu/abs/2007MNRAS.377.1043M} {377, 1043}

\bibitem[\protect\citeauthoryear{{Mesinger}}{{Mesinger}}{2010}]{mesinger2010}
{Mesinger} A.,  2010, \mn@doi [\mnras] {10.1111/j.1365-2966.2010.16995.x},
  \href {https://ui.adsabs.harvard.edu/abs/2010MNRAS.407.1328M} {407, 1328}

\bibitem[\protect\citeauthoryear{Mesinger}{Mesinger}{2016}]{mesinger2016a}
Mesinger A.,  2016, Understanding the Epoch of Cosmic Reionization: Challenges
  and Progress.
 Vol. 423, Springer, \mn@doi{10.1007/978-3-319-21957-8}

\bibitem[\protect\citeauthoryear{{Mesinger} \& {Furlanetto}}{{Mesinger} \&
  {Furlanetto}}{2007}]{mesinger2007}
{Mesinger} A.,  {Furlanetto} S.,  2007, \mn@doi [\apj] {10.1086/521806}, \href
  {https://ui.adsabs.harvard.edu/abs/2007ApJ...669..663M} {669, 663}

\bibitem[\protect\citeauthoryear{{Mesinger}, {Furlanetto}  \& {Cen}}{{Mesinger}
  et~al.}{2011}]{mesinger2011}
{Mesinger} A.,  {Furlanetto} S.,   {Cen} R.,  2011, \mn@doi [\mnras]
  {10.1111/j.1365-2966.2010.17731.x}, \href
  {https://ui.adsabs.harvard.edu/abs/2011MNRAS.411..955M} {411, 955}

\bibitem[\protect\citeauthoryear{{Mesinger}, {Aykutalp}, {Vanzella},
  {Pentericci}, {Ferrara}  \& {Dijkstra}}{{Mesinger}
  et~al.}{2015}]{mesinger2015}
{Mesinger} A.,  {Aykutalp} A.,  {Vanzella} E.,  {Pentericci} L.,  {Ferrara} A.,
    {Dijkstra} M.,  2015, \mn@doi [\mnras] {10.1093/mnras/stu2089}, \href
  {https://ui.adsabs.harvard.edu/abs/2015MNRAS.446..566M} {446, 566}

\bibitem[\protect\citeauthoryear{{Mesinger}, {Greig}  \& {Sobacchi}}{{Mesinger}
  et~al.}{2016}]{mesinger2016b}
{Mesinger} A.,  {Greig} B.,   {Sobacchi} E.,  2016, \mn@doi [\mnras]
  {10.1093/mnras/stw831}, \href
  {https://ui.adsabs.harvard.edu/abs/2016MNRAS.459.2342M} {459, 2342}

\bibitem[\protect\citeauthoryear{{More}, {van den Bosch}, {Cacciato}, {Mo},
  {Yang}  \& {Li}}{{More} et~al.}{2009}]{more2009}
{More} S.,  {van den Bosch} F.~C.,  {Cacciato} M.,  {Mo} H.~J.,  {Yang} X.,
  {Li} R.,  2009, \mn@doi [\mnras] {10.1111/j.1365-2966.2008.14095.x}, \href
  {https://ui.adsabs.harvard.edu/abs/2009MNRAS.392..801M} {392, 801}

\bibitem[\protect\citeauthoryear{{Moster}, {Naab}  \& {White}}{{Moster}
  et~al.}{2013}]{moster2013}
{Moster} B.~P.,  {Naab} T.,   {White} S. D.~M.,  2013, \mn@doi [\mnras]
  {10.1093/mnras/sts261}, \href
  {https://ui.adsabs.harvard.edu/abs/2013MNRAS.428.3121M} {428, 3121}

\bibitem[\protect\citeauthoryear{{Murray}, {Power}  \& {Robotham}}{{Murray}
  et~al.}{2013}]{murray2013}
{Murray} S.~G.,  {Power} C.,   {Robotham} A.~S.~G.,  2013, \mn@doi [Astronomy
  and Computing] {10.1016/j.ascom.2013.11.001}, \href
  {https://ui.adsabs.harvard.edu/abs/2013A&C.....3...23M} {3, 23}

\bibitem[\protect\citeauthoryear{{Neufeld}}{{Neufeld}}{1990}]{neufeld1990}
{Neufeld} D.~A.,  1990, \mn@doi [\apj] {10.1086/168375}, \href
  {https://ui.adsabs.harvard.edu/abs/1990ApJ...350..216N} {350, 216}

\bibitem[\protect\citeauthoryear{{Oesch} et~al.,}{{Oesch}
  et~al.}{2015}]{oesch2015}
{Oesch} P.~A.,  et~al., 2015, \mn@doi [\apjl] {10.1088/2041-8205/804/2/L30},
  \href {https://ui.adsabs.harvard.edu/abs/2015ApJ...804L..30O} {804, L30}

\bibitem[\protect\citeauthoryear{Oliphant}{Oliphant}{2006}]{oliphant2006}
Oliphant T.~E.,  2006, A guide to NumPy.
 Vol. 1, Trelgol Publishing USA

\bibitem[\protect\citeauthoryear{{Ono} et~al.,}{{Ono} et~al.}{2012}]{ono2012}
{Ono} Y.,  et~al., 2012, \mn@doi [\apj] {10.1088/0004-637X/744/2/83}, \href
  {https://ui.adsabs.harvard.edu/abs/2012ApJ...744...83O} {744, 83}

\bibitem[\protect\citeauthoryear{{Ouchi} et~al.,}{{Ouchi}
  et~al.}{2010}]{ouchi2010}
{Ouchi} M.,  et~al., 2010, \mn@doi [\apj] {10.1088/0004-637X/723/1/869}, \href
  {https://ui.adsabs.harvard.edu/abs/2010ApJ...723..869O} {723, 869}

\bibitem[\protect\citeauthoryear{{Ouchi} et~al.,}{{Ouchi}
  et~al.}{2018}]{ouchi2018}
{Ouchi} M.,  et~al., 2018, \mn@doi [\pasj] {10.1093/pasj/psx074}, \href
  {https://ui.adsabs.harvard.edu/abs/2018PASJ...70S..13O} {70, S13}

\bibitem[\protect\citeauthoryear{{Paardekooper}, {Khochfar}  \& {Dalla
  Vecchia}}{{Paardekooper} et~al.}{2015}]{pardekooper2015}
{Paardekooper} J.-P.,  {Khochfar} S.,   {Dalla Vecchia} C.,  2015, \mn@doi
  [\mnras] {10.1093/mnras/stv1114}, \href
  {https://ui.adsabs.harvard.edu/abs/2015MNRAS.451.2544P} {451, 2544}

\bibitem[\protect\citeauthoryear{{Parsa}, {Dunlop}  \& {McLure}}{{Parsa}
  et~al.}{2018}]{parsa2018}
{Parsa} S.,  {Dunlop} J.~S.,   {McLure} R.~J.,  2018, \mn@doi [\mnras]
  {10.1093/mnras/stx2887}, \href
  {https://ui.adsabs.harvard.edu/abs/2018MNRAS.474.2904P} {474, 2904}

\bibitem[\protect\citeauthoryear{{Parzen}}{{Parzen}}{1962}]{parzen1962}
{Parzen} E.,  1962, \mn@doi [Ann. Math. Statist.] {10.1214/aoms/1177704472},
  33, 1065

\bibitem[\protect\citeauthoryear{{Pentericci} et~al.,}{{Pentericci}
  et~al.}{2011}]{pentericci2011}
{Pentericci} L.,  et~al., 2011, \mn@doi [\apj] {10.1088/0004-637X/743/2/132},
  \href {https://ui.adsabs.harvard.edu/abs/2011ApJ...743..132P} {743, 132}

\bibitem[\protect\citeauthoryear{{Pentericci} et~al.,}{{Pentericci}
  et~al.}{2014}]{pentericci2014}
{Pentericci} L.,  et~al., 2014, \mn@doi [\apj] {10.1088/0004-637X/793/2/113},
  \href {https://ui.adsabs.harvard.edu/abs/2014ApJ...793..113P} {793, 113}

\bibitem[\protect\citeauthoryear{{Perez} \& {Granger}}{{Perez} \&
  {Granger}}{2007}]{perez2007}
{Perez} F.,  {Granger} B.~E.,  2007, \mn@doi [Computing in Science and
  Engineering] {10.1109/MCSE.2007.53}, \href
  {https://ui.adsabs.harvard.edu/abs/2007CSE.....9c..21P} {9, 21}

\bibitem[\protect\citeauthoryear{{Planck Collaboration} et~al.,}{{Planck
  Collaboration} et~al.}{2016a}]{planck2016a}
{Planck Collaboration} et~al., 2016a, \mn@doi [\aap]
  {10.1051/0004-6361/201525830}, \href
  {https://ui.adsabs.harvard.edu/abs/2016A&A...594A..13P} {594, A13}

\bibitem[\protect\citeauthoryear{{Planck Collaboration} et~al.,}{{Planck
  Collaboration} et~al.}{2016b}]{planck2016b}
{Planck Collaboration} et~al., 2016b, \mn@doi [\aap]
  {10.1051/0004-6361/201628897}, \href
  {https://ui.adsabs.harvard.edu/abs/2016A&A...596A.108P} {596, A108}

\bibitem[\protect\citeauthoryear{{Planck Collaboration} et~al.,}{{Planck
  Collaboration} et~al.}{2018}]{planck2018}
{Planck Collaboration} et~al., 2018, arXiv e-prints, \href
  {https://ui.adsabs.harvard.edu/abs/2018arXiv180706209P} {p. arXiv:1807.06209}

\bibitem[\protect\citeauthoryear{{Ren}, {Trenti}  \& {Mutch}}{{Ren}
  et~al.}{2018}]{ren2018}
{Ren} K.,  {Trenti} M.,   {Mutch} S.~J.,  2018, \mn@doi [\apj]
  {10.3847/1538-4357/aab094}, \href
  {https://ui.adsabs.harvard.edu/abs/2018ApJ...856...81R} {856, 81}

\bibitem[\protect\citeauthoryear{{Ren}, {Trenti}  \& {Mason}}{{Ren}
  et~al.}{2019}]{ren2019}
{Ren} K.,  {Trenti} M.,   {Mason} C.~A.,  2019, \mn@doi [\apj]
  {10.3847/1538-4357/ab2117}, \href
  {https://ui.adsabs.harvard.edu/abs/2019ApJ...878..114R} {878, 114}

\bibitem[\protect\citeauthoryear{{Roberts-Borsani} et~al.,}{{Roberts-Borsani}
  et~al.}{2016}]{roberts-borsani2016}
{Roberts-Borsani} G.~W.,  et~al., 2016, \mn@doi [\apj]
  {10.3847/0004-637X/823/2/143}, \href
  {https://ui.adsabs.harvard.edu/abs/2016ApJ...823..143R} {823, 143}

\bibitem[\protect\citeauthoryear{{Robertson}, {Ellis}, {Dunlop}, {McLure}  \&
  {Stark}}{{Robertson} et~al.}{2010}]{robertson2010}
{Robertson} B.~E.,  {Ellis} R.~S.,  {Dunlop} J.~S.,  {McLure} R.~J.,   {Stark}
  D.~P.,  2010, \mn@doi [\nat] {10.1038/nature09527}, \href
  {https://ui.adsabs.harvard.edu/abs/2010Natur.468...49R} {468, 49}

\bibitem[\protect\citeauthoryear{{Rosenblatt}}{{Rosenblatt}}{1956}]{rosenblatt1956}
{Rosenblatt} M.,  1956, \mn@doi [Proceedings of the National Academy of
  Science] {10.1073/pnas.42.1.43}, \href
  {https://ui.adsabs.harvard.edu/abs/1956PNAS...42...43R} {42, 43}

\bibitem[\protect\citeauthoryear{{Santos}}{{Santos}}{2004}]{santos2004}
{Santos} M.~R.,  2004, \mn@doi [\mnras] {10.1111/j.1365-2966.2004.07594.x},
  \href {https://ui.adsabs.harvard.edu/abs/2004MNRAS.349.1137S} {349, 1137}

\bibitem[\protect\citeauthoryear{{Schenker}, {Stark}, {Ellis}, {Robertson},
  {Dunlop}, {McLure}, {Kneib}  \& {Richard}}{{Schenker}
  et~al.}{2012}]{schenker2012}
{Schenker} M.~A.,  {Stark} D.~P.,  {Ellis} R.~S.,  {Robertson} B.~E.,  {Dunlop}
  J.~S.,  {McLure} R.~J.,  {Kneib} J.-P.,   {Richard} J.,  2012, \mn@doi [\apj]
  {10.1088/0004-637X/744/2/179}, \href
  {https://ui.adsabs.harvard.edu/abs/2012ApJ...744..179S} {744, 179}

\bibitem[\protect\citeauthoryear{{Sheth}, {Mo}  \& {Tormen}}{{Sheth}
  et~al.}{2001}]{sheth2001}
{Sheth} R.~K.,  {Mo} H.~J.,   {Tormen} G.,  2001, \mn@doi [\mnras]
  {10.1046/j.1365-8711.2001.04006.x}, \href
  {https://ui.adsabs.harvard.edu/abs/2001MNRAS.323....1S} {323, 1}

\bibitem[\protect\citeauthoryear{{Shibuya} et~al.,}{{Shibuya}
  et~al.}{2014}]{shibuya2014}
{Shibuya} T.,  et~al., 2014, \mn@doi [\apj] {10.1088/0004-637X/788/1/74}, \href
  {https://ui.adsabs.harvard.edu/abs/2014ApJ...788...74S} {788, 74}

\bibitem[\protect\citeauthoryear{{Sobacchi} \& {Mesinger}}{{Sobacchi} \&
  {Mesinger}}{2014}]{sobacchi2014}
{Sobacchi} E.,  {Mesinger} A.,  2014, \mn@doi [\mnras] {10.1093/mnras/stu377},
  \href {https://ui.adsabs.harvard.edu/abs/2014MNRAS.440.1662S} {440, 1662}

\bibitem[\protect\citeauthoryear{{Sobacchi} \& {Mesinger}}{{Sobacchi} \&
  {Mesinger}}{2015}]{sobacchi2015}
{Sobacchi} E.,  {Mesinger} A.,  2015, \mn@doi [\mnras] {10.1093/mnras/stv1751},
  \href {https://ui.adsabs.harvard.edu/abs/2015MNRAS.453.1843S} {453, 1843}

\bibitem[\protect\citeauthoryear{{Sobacchi}, {Mesinger}  \& {Greig}}{{Sobacchi}
  et~al.}{2016}]{sobacchi2016}
{Sobacchi} E.,  {Mesinger} A.,   {Greig} B.,  2016, \mn@doi [\mnras]
  {10.1093/mnras/stw811}, \href
  {https://ui.adsabs.harvard.edu/abs/2016MNRAS.459.2741S} {459, 2741}

\bibitem[\protect\citeauthoryear{{Spergel} et~al.,}{{Spergel}
  et~al.}{2015}]{spergel2015}
{Spergel} D.,  et~al., 2015, arXiv e-prints, \href
  {https://ui.adsabs.harvard.edu/abs/2015arXiv150303757S} {p. arXiv:1503.03757}

\bibitem[\protect\citeauthoryear{{Stark}, {Ellis}, {Chiu}, {Ouchi}  \&
  {Bunker}}{{Stark} et~al.}{2010}]{stark2010}
{Stark} D.~P.,  {Ellis} R.~S.,  {Chiu} K.,  {Ouchi} M.,   {Bunker} A.,  2010,
  \mn@doi [\mnras] {10.1111/j.1365-2966.2010.17227.x}, \href
  {https://ui.adsabs.harvard.edu/abs/2010MNRAS.408.1628S} {408, 1628}

\bibitem[\protect\citeauthoryear{{Stark} et~al.,}{{Stark}
  et~al.}{2017}]{stark2017}
{Stark} D.~P.,  et~al., 2017, \mn@doi [\mnras] {10.1093/mnras/stw2233}, \href
  {https://ui.adsabs.harvard.edu/abs/2017MNRAS.464..469S} {464, 469}

\bibitem[\protect\citeauthoryear{{Steidel}, {Erb}, {Shapley}, {Pettini},
  {Reddy}, {Bogosavljevi{\'c}}, {Rudie}  \& {Rakic}}{{Steidel}
  et~al.}{2010}]{steidel2010}
{Steidel} C.~C.,  {Erb} D.~K.,  {Shapley} A.~E.,  {Pettini} M.,  {Reddy} N.,
  {Bogosavljevi{\'c}} M.,  {Rudie} G.~C.,   {Rakic} O.,  2010, \mn@doi [\apj]
  {10.1088/0004-637X/717/1/289}, \href
  {https://ui.adsabs.harvard.edu/abs/2010ApJ...717..289S} {717, 289}

\bibitem[\protect\citeauthoryear{{Tacchella}, {Bose}, {Conroy}, {Eisenstein}
  \& {Johnson}}{{Tacchella} et~al.}{2018}]{tacchella2018}
{Tacchella} S.,  {Bose} S.,  {Conroy} C.,  {Eisenstein} D.~J.,   {Johnson}
  B.~D.,  2018, \mn@doi [\apj] {10.3847/1538-4357/aae8e0}, \href
  {https://ui.adsabs.harvard.edu/abs/2018ApJ...868...92T} {868, 92}

\bibitem[\protect\citeauthoryear{{Treu}, {Trenti}, {Stiavelli}, {Auger}  \&
  {Bradley}}{{Treu} et~al.}{2012}]{treu2012}
{Treu} T.,  {Trenti} M.,  {Stiavelli} M.,  {Auger} M.~W.,   {Bradley} L.~D.,
  2012, \mn@doi [\apj] {10.1088/0004-637X/747/1/27}, \href
  {https://ui.adsabs.harvard.edu/abs/2012ApJ...747...27T} {747, 27}

\bibitem[\protect\citeauthoryear{{Verhamme}, {Schaerer}  \&
  {Maselli}}{{Verhamme} et~al.}{2006}]{verhamme2006}
{Verhamme} A.,  {Schaerer} D.,   {Maselli} A.,  2006, \mn@doi [\aap]
  {10.1051/0004-6361:20065554}, \href
  {https://ui.adsabs.harvard.edu/abs/2006A&A...460..397V} {460, 397}

\bibitem[\protect\citeauthoryear{{Verhamme}, {Schaerer}, {Atek}  \&
  {Tapken}}{{Verhamme} et~al.}{2008}]{verhamme2008}
{Verhamme} A.,  {Schaerer} D.,  {Atek} H.,   {Tapken} C.,  2008, \mn@doi [\aap]
  {10.1051/0004-6361:200809648}, \href
  {https://ui.adsabs.harvard.edu/abs/2008A&A...491...89V} {491, 89}

\bibitem[\protect\citeauthoryear{{Weinberger}, {Kulkarni}, {Haehnelt},
  {Choudhury}  \& {Puchwein}}{{Weinberger} et~al.}{2018}]{weinberger2018}
{Weinberger} L.~H.,  {Kulkarni} G.,  {Haehnelt} M.~G.,  {Choudhury} T.~R.,
  {Puchwein} E.,  2018, \mn@doi [\mnras] {10.1093/mnras/stu979}, \href
  {https://ui.adsabs.harvard.edu/abs/2014MNRAS.442.2560W} {479, 2560}

\bibitem[\protect\citeauthoryear{{Weinberger}, {Haehnelt}  \&
  {Kulkarni}}{{Weinberger} et~al.}{2019}]{weinberger2019}
{Weinberger} L.~H.,  {Haehnelt} M.~G.,   {Kulkarni} G.,  2019, \mn@doi [\mnras]
  {10.1093/mnras/stz481}, \href
  {https://ui.adsabs.harvard.edu/abs/2019MNRAS.485.1350W} {485, 1350}

\bibitem[\protect\citeauthoryear{{Wise}, {Demchenko}, {Halicek}, {Norman},
  {Turk}, {Abel}  \& {Smith}}{{Wise} et~al.}{2014}]{wise2014}
{Wise} J.~H.,  {Demchenko} V.~G.,  {Halicek} M.~T.,  {Norman} M.~L.,  {Turk}
  M.~J.,  {Abel} T.,   {Smith} B.~D.,  2014, \mn@doi [\mnras]
  {10.1093/mnras/stu979}, \href
  {https://ui.adsabs.harvard.edu/abs/2014MNRAS.442.2560W} {442, 2560}

\bibitem[\protect\citeauthoryear{{Yang}, {Mo}, {Jing}  \& {van den
  Bosch}}{{Yang} et~al.}{2005}]{yang2005}
{Yang} X.,  {Mo} H.~J.,  {Jing} Y.~P.,   {van den Bosch} F.~C.,  2005, \mn@doi
  [\mnras] {10.1111/j.1365-2966.2005.08801.x}, \href
  {https://ui.adsabs.harvard.edu/abs/2005MNRAS.358..217Y} {358, 217}

\bibitem[\protect\citeauthoryear{{Yang} et~al.,}{{Yang}
  et~al.}{2017}]{yang2017}
{Yang} H.,  et~al., 2017, \mn@doi [\apj] {10.3847/1538-4357/aa7d4d}, \href
  {https://ui.adsabs.harvard.edu/abs/2017ApJ...844..171Y} {844, 171}

\bibitem[\protect\citeauthoryear{{Zitrin} et~al.,}{{Zitrin}
  et~al.}{2015}]{zitrin2015}
{Zitrin} A.,  et~al., 2015, \mn@doi [\apjl] {10.1088/2041-8205/810/1/L12},
  \href {https://ui.adsabs.harvard.edu/abs/2015ApJ...810L..12Z} {810, L12}

\bibitem[\protect\citeauthoryear{{van der Walt}, {Colbert}  \&
  {Varoquaux}}{{van der Walt} et~al.}{2011}]{vanderWalt2011}
{van der Walt} S.,  {Colbert} S.~C.,   {Varoquaux} G.,  2011, \mn@doi
  [Computing in Science and Engineering] {10.1109/MCSE.2011.37}, \href
  {https://ui.adsabs.harvard.edu/abs/2011CSE....13b..22V} {13, 22}

\makeatother
\end{thebibliography}


\appendix

\section{Comparison of Biases in the Neutral Fraction Inference} \label{appendix:sim_inference}

\begin{figure*}
    \centering
	\includegraphics[width=\textwidth]{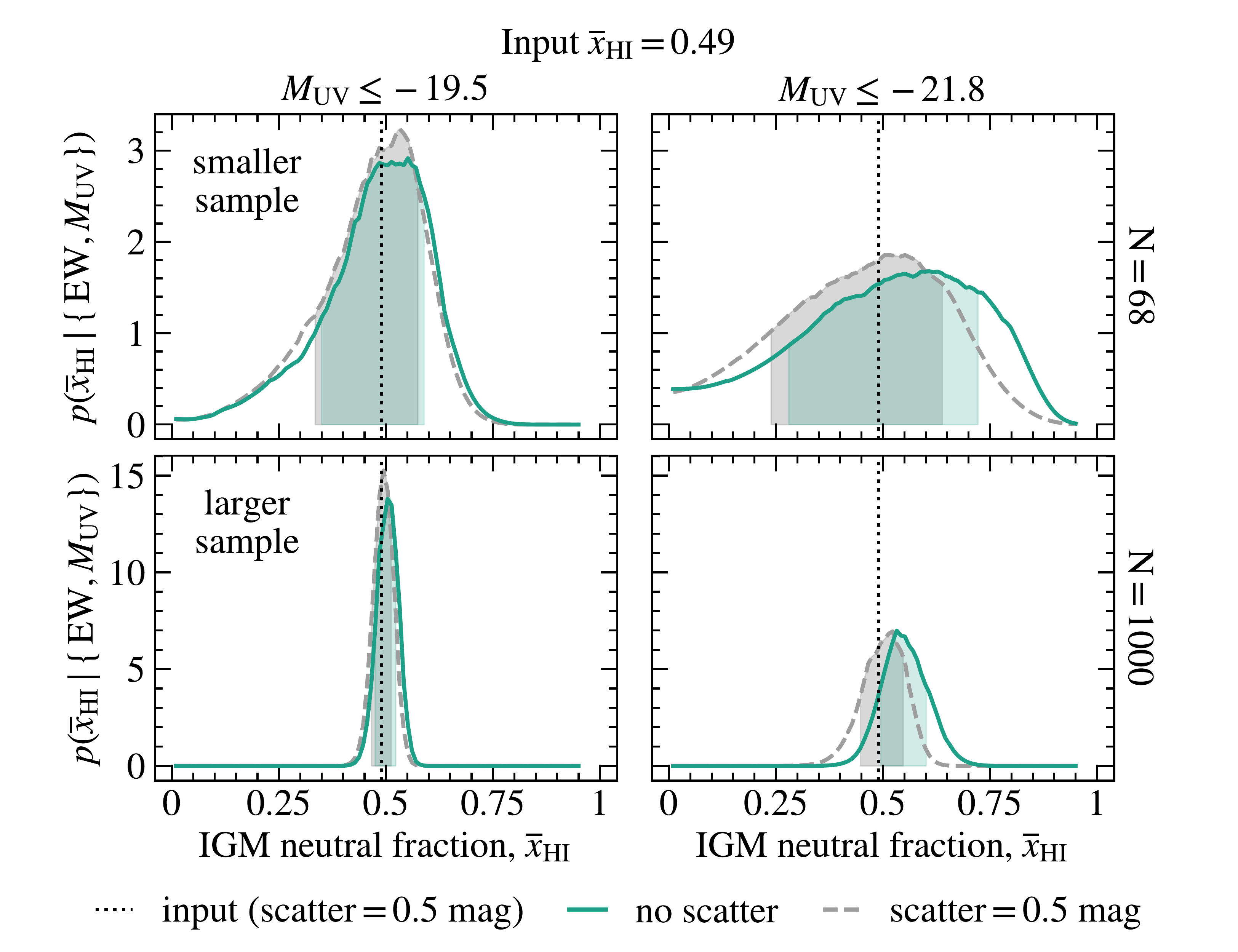}
	\caption{Posterior distributions for \xHI\ inferred from four simulated samples of galaxies with various sample sizes and limiting absolute magnitudes with (dashed grey) and without (solid green) accounting for scatter in the UV luminosity--halo mass relation. Columns correspond to samples with limiting magnitudes of $M_\textsc{uv} \leq -19.5$ (left) and $M_\textsc{uv} \leq -21.8$ (right), and rows correspond to sample sizes of $N = 68$ (top) and $N = 1000$ (bottom). Black dotted lines denote the input neutral fraction ($\overline{x}_\ion{H}{i} = 0.49$) and the shaded regions are the 68 per cent credible intervals. With the smaller sample, the inferences both with and without scatter recover the input neutral fraction within errors. However, for the larger samples, the posterior becomes more precise, and the inference with limiting magnitude $M_\textsc{uv} = -21.8$ without scatter overestimates the neutral fraction. Thus, accounting for UV luminosity--halo mass relation scatter becomes important for large samples of bright galaxies.}
	\label{fig:sim_comparison}
\end{figure*}

We compare the inferred neutral fraction with and without accounting for scatter for four samples of simulated galaxies: two samples ($N = 68$ and $N = 1000$ galaxies) biased towards bright ($M_\textsc{uv} \leq -21.8$) galaxies, and two fainter samples with $M_\textsc{uv} \leq -19.5$. We draw galaxies from the \citet{mason2015} UV luminosity function at \z{7} and populate them with \Lya\ EWs from the probability distribution $p(EW \mid M_\textsc{uv}, \overline{x}_\ion{H}{i}, \sigma)$ with scatter of $\sigma = 0.5$\,mag at a neutral fraction of $\overline{x}_\ion{H}{i} = 0.49$. We furthermore add Gaussian noise with a standard deviation of 5\,\AA\ to the \Lya\ EWs. We then infer the neutral fraction with and without accounting for scatter in the UV luminosity--halo mass relation.

Fig.\ \ref{fig:sim_comparison} shows the posterior distributions for \xHI\ inferred with and without accounting for scatter in the UV luminosity--halo mass relation for all four samples. For the smaller sample size of $N = 68$, the inferences both with and without scatter recover the input neutral fraction of $\overline{x}_\ion{H}{i} = 0.49$ within the uncertainties. However, for the larger sample of $N = 1000$, the inference without scatter performs more poorly as the posteriors become more precise: the inference without scatter for the samples of galaxies with $M_\textsc{uv} \leq -21.8$ systematically overestimates the input neutral fraction, which is more significant for the larger sample size. In contrast, the inference without scatter using the sample with $M_\textsc{uv} \leq -19.5$ recovers the neutral fraction well for both sample sizes. This is because scatter in the UV luminosity--halo mass relation primarily impacts bright galaxies, but fainter galaxies are much more numerous and thus contribute significantly to the inference when included. Thus, the effects of scatter on UV-bright galaxies do not bias neutral fraction inferences when the observed sample include galaxies $M_\textsc{uv} > M_\textsc{uv}^\star$.

To further assess the impact of including scatter in the UV luminosity--halo mass relation, we evaluate the Kullback-Leibler divergence \citep[KL divergence;][]{kullback1968} between the posterior distributions with and without accounting for scatter for each sample of simulated galaxies. The KL divergence between the posteriors with and without scatter is larger for the sample with the brighter magnitude limit, reflecting the increased impact of scatter on \Lya\ transmission through the reionizing IGM for brighter galaxies, which biases the inferred neutral fraction. The KL divergence between the two posteriors increases with increasing sample size, reflecting the more precise constraints on the neutral fraction obtained from larger samples. 


\bsp	
\label{lastpage}
\end{document}